# Drying of soft colloidal films

*Keumkyung Kuk,[a] Julian Ringling,[a] Kevin Gräff,[b] Sebastian Hänsch,[c] Virginia Carrasco-Fadanelli,[d] Andrey A. Rudov,[e,f] Igor I. Potemkin,[e,f] Regine von Klitzing,[b] Ivo Buttinoni,[d] and Matthias Karg[a]\**

[a]Institut für Physikalische Chemie I: Kolloide und Nanooptik, Heinrich-Heine-Universität Düsseldorf, Universitätsstr. 1, 40225 Düsseldorf, Germany

[b]Institute for Condensed Matter Physics, Soft Matter at Interfaces, Technische Universität Darmstadt, Hochschulstr. 8, 64289 Darmstadt, Germany

[c]Center for Advanced Imaging, Heinrich-Heine-Universität Düsseldorf, Universitätsstr. 1, 40225 Düsseldorf, Germany

[d]Institut für Experimentelle Physik der kondensierten Materie, Heinrich-Heine-Universität Düsseldorf, Universitätsstr. 1, 40225 Düsseldorf, Germany

[e]DWI-Leibniz Institute for Interactive Materials, 52056 Aachen, Germany

[f]Physics Department, Lomonosov Moscow State University, Moscow 119991, Russian Federation

## Abstract

Thin films made of deformable micro- and nano-units, such as biological membranes, polymer interfaces, and particle-laden liquid surfaces, exhibit a complex behavior during drying, with consequences for various applications like wound healing, coating technologies, and additive manufacturing. Studying the drying dynamics and structural changes of soft colloidal films thus holds the potential to yield valuable insights to achieve improvements for applications. In this study, we employ interfacial monolayers of core-shell microgels with varying degrees of softness as model systems and investigate their drying behavior on differently modified solid substrates (hydrophobic vs. hydrophilic). By leveraging on video microscopy, particle tracking, and thin film interference, we shed light on the interplay between microgel adhesion to solid surfaces and the immersion capillary forces that arise in the thin liquid film. We discovered that a dried replica of the interfacial microstructure can be more accurately achieved on a hydrophobic substrate relative to a hydrophilic one, particularly when employing softer colloids as opposed to harder counterparts. These observations are qualitatively supported by experiments with a thin film pressure balance which allows mimicking and controlling the drying process and by computer simulations with coarse-grained models.

## 1 Introduction

Surface-active agents are ubiquitous in our daily lives and paramount for numerous industrial applications including emulsion stabilization, foam formation, detergency, pharmaceutical drug delivery, and structured surfaces.[1] They often consist of amphiphilic molecules (surfactants) that adsorb onto the interface between two immiscible phases (e.g., water and air), lower the surface tension, and stabilize the interface, e.g., against droplet coalescence. Remarkably, the same stabilization can be achieved via adsorption of solid micro- and nanoparticles, as in the case of Pickering emulsions.[2] Compared to molecular surfactants, solid particles are believed to offer enhanced stability over extended periods of time, potentially exhibit lower toxicity,[3] and generate interfacial films with significantly greater thickness due to their larger size.[4] Additionally, solid particles need longer times to adsorb (in line with their characteristic diffusion timescale). Their presence at the interfaces often leads to the deformation of the interface due to their weight/size,[5] shape,[6] porosity,[7] and/or wetting properties,[8] which give rise to capillary interactions among adjacent particles. For example, the attractive force between two particles at the air/water interface (flotation capillary force) increases with the sixth power of the particle size and becomes negligible when the particle is smaller than 10 μm.[9-10] Conversely, particles as small as protein globules can be subjected to the influence of a strong attractive force when trapped in a thin water film (immersion capillary force).[11] The horizontal projection of the attractive (immersion) capillary force $F_x$ between two (solid) spheres of radius $R$ at a center-to-center distance $D_{c-c}$ partially immersed in a liquid is given by:[12]

$$F_x \approx 2\pi\sigma R_c^2 (\sin^2 \Psi_c) \left(\frac{1}{D_{c-c}}\right) \qquad (1)$$

Here, σ is the surface tension, $R_c$ the radius of the three-phase contact line at the particle surface and $\Psi_c$ is the mean meniscus slope angle.[11] Even for submicrometer sized particles, the capillary attraction can be much larger than the thermal energy. Since the interaction force is directly proportional to σ, capillary attraction is particularly strong in drying of aqueous films.

Interfacial deformations, that ultimately also dictate $\Psi_c$, have a dramatic impact on the way a film dries on rigid surfaces, potentially leading to highly inhomogeneous drying conditions during coating and deposition processes.[13-15] Inhomogeneous drying is known from the segregation of solutes to the edge of drying droplets leading to a ring-like deposit ("coffe ring effect").[16] This phenomenon is driven by outward fluid flow caused by non-uniform evaporation flux.[16] On the other hand a gradient in interfacial tension in a liquid film leads to flow caused by the Marangoni effect.[17] Understanding of such effects that strongly influence the drying of colloidal films is crucial in many areas including printing, colloidal self-assembly and coating technologies. The problem of inhomogeneous drying can be potentially solved by introducing strong repulsion among the solid particles[17-18] or employing "sticky" and deformable polymeric chains/networks.[19-20] Examples are poly(*N*-isopropylacrylamide) (PNIPAM) microgels, i.e. soft colloid-like objects with an internal gel-like structure composed of a highly swollen crosslinked polymer network.[21] Through simple precipitation polymerization, such microgels can be synthesized with some control over their internal architecture, size, and softness. Much like their solid counterparts, PNIPAM-based microgels also spontaneously adsorb at various interfaces, e.g., foams, foam films,[22] and emulsions.[23-26] At these fluid interfaces microgels are stretched along the interfacial plane, assuming a "fried egg" shape.[27-28] At large enough packing fractions, microgels can be finally assembled into closed-packed viscoelastic films of high porosity−properties that are reminiscent of biological interactions (e.g., cell adhesion[29-30]) and interfaces.[31] Such close-packed microgel films can be used to prepare thermoresponsive, free-standing membranes that allow for manipulating ion transport by temperature.[32] Controlling the packing fraction of microgels at fluid interfaces has also been used to prepare ordered monolayers on rigid substrates that can be used to create complex tesselations via self-templating assembly.[33] Core-shell microgels with plasmonic gold or silver nanoparticle cores can be used to fabricate plasmonic surface coatings.[34-35] The optical properties of such structures crucially depend on packing fraction and the structure factor of the assembly.[36-38] All of these examples rely on drying of thin liquid films and the studied properties are strongly correlated to the homogeneity of the resulting microstructure.

To date, the microstructure and mechanical properties of microgel-laden interfaces have been widely studied using Langmuir trough setups, which offer an excellent experimental platform with a flat fluid interface and controlled conditions. These setups involve fluid interface-assisted assembly by means of lateral compression, often followed by deposition onto solid substrates (Langmuir-Blodgett deposition). While studies on microgels with various sizes, charge, and crosslinker densities (softness) have yielded valuable insights into the behavior of soft colloids at different fluid interfaces,[39-41] very little is known about how these films dry,[37] even though drying effects could significantly alter the microstructure of a microgel-laden fluid interface upon transfer onto solid substrates.[42-43] In particular, to our knowledge, there has not yet been a study that examines *in situ* drying dynamics at the single-microgel level. The research gap exists primarily because visualizing interfaces deformed by soft micro-units is difficult, and a comprehensive theoretical framework that describes the complex physics during drying is yet to be established.[44-47]

In this work, we use a combination of video microscopy and thin film interference to study the drying process of microgel interfacial films on differently modified substrates (hydrophobic vs. hydrophilic). By considering micron-sized silica-PNIPAM core-shell (CS) microgels[48] of comparable sizes and different crosslinker densities (i.e., softness), we trace the drying dynamics at the single microgel level and elucidate the interplay between two key elements: the microgel-to-substrate adhesion and the capillary forces experienced by microgels in an immersed state (immersion capillary force). Relatively homogeneous drying is reported for films containing microgels with a low crosslinker density drying on hydrophobic substrates, whereas significant alterations of the original microstructures occur when highly crosslinked microgels are deposited on hydrophilic solid surfaces. Similar effects during the deposition of other soft films (e.g., lipid-laden interfaces) have been observed.[49] Understanding the behavior of such soft colloids in thin films not only enhances fundamental knowledge but also has practical implications across various industries, such as microfluidics, surface nanopatterning, biomedical devices, food, cosmetics, and coating.

## 2 Results and Discussion

### 2.1 Experimental design

Predicting how soft interfacial films dry onto solid substrates is far from trivial. In our model system consisting of microgel-laden interfaces transferred onto solid substrates, a rich drying scenario emerged depending on (1) the surface chemistry of the substrate, (2) the softness of the individual microgels, and (3) the surface pressure of the microgel-laden interface. We adjusted these parameters by (1) functionalizing the solid surface, (2) tuning the amount of crosslinker during microgel synthesis and (3) compressing the colloidal films at the fluid interface.

To achieve high optical contrast we synthesized three batches of CS microgels that are composed of monodisperse, rigid cores, and PNIPAM hydrogel shells of different crosslinker densities: CS-low (1.0 mol.% crosslinker by design), CS-medium (2.5 mol.%), and CS-high (7.5 mol.%). The structure of these microgels is schematiclly illustrated in **Figure 1**. These three model microgels share the same batch of cores (silica, diameter $D_c$ = 437 nm) and have a similar overall hydrodynamic diameter, $D_h$, of about 1 μm (measured by dynamic light scattering at 20 °C). For comparison, "classical", purely organic microgels without solid cores were also synthesized and investigated. More details about the preparation and characterization can be found in the **Experimental Section**. The variation in crosslinker density of PNIPAM-based microgels is clearly observed in the temperature-dependent hydrodynamic diameter and compression isotherms, as previously reported by several studies.[41, 50-53] This difference in crosslinker density among the microgels also becomes apparent through the degree of lateral deformation at the air/water interface, driven by interfacial tension. In this study, we use the term "softness" to describe the deformation behavior of model microgels across diverse scenarios, with CS-low being the softest of all and CS-high the stiffest. However, it is important to note that strictly, at interfaces, we deal with a non-isotropic softness within the polymeric network.[54] Additionally, the crosslinker density is not homogeneous within the shell, being higher near the core and lower in the outer layer. Despite these complexities, the primary determinant of microgel "softness" is known to be the crosslinker density.[52]

When CS microgels adsorb at the air/water interface, they tend to cluster, likely due to their irregular meniscus. This lateral attraction, commonly observed in colloids at fluid interfaces as a result of meniscus undulations, can be described using a multipole expansion. The leading quadrupolar term in this expansion produces a net attractive force.[9-11] The depth of the potential well (where the capillary attraction overcomes the electrostatic repulsion), is calculated to be $10^4$ $k_B T$ for a solid particle with 1 μm diameter and 50 nm deviations in the meniscus,[55] where $k_B$ is the Boltzmann constant. However, this attraction is reported to be reduced for porous materials (such as microgels) by an order of magnitude.[7] The degree of lateral deformation in dependence of softness is quantified by the ratio between the interfacial diameter at the air/water interface, $D_i$, normalized by $D_h$. We define $D_i$ as the mean nearest center-to-center distance, $D_{c-c}$, at surface pressure $\Pi \approx 0$ mN/m, under the condition that the microgels are clustered, i.e., already in contact.

The values of $D_i/D_h$ are shown in **Figure 1** and listed in the **Supporting Information (SI)** in **Table S1**. The $D_i$ measurements were done by reflected light microscopy with 30 minutes of equilibration time after the adsorption of CS microgels at the air/water interface from ethanolic dispersion (ethanol as the spreading agent) as well as from aqueous dispersion (spontaneous adsorption from subphase), where they form clusters, a precondition for our definition of $D_i$. These clusters of CS microgels and their corresponding compression isotherms are shown in **Figure S1**. While the CS-low microgels assume approximately 1.7 times their (lateral) diameter at the fluid interface as compared to $D_h$, the stiffest microgels, CS-high, show only little deformation at the interface ($D_i/D_h \approx 1$). The use of ethanol as spreading agent as compared to spontaneuous adsorption from the purely aqueous subphase has very little effect on the degree of deformation at the interface. The degree of deformation for CS-low (1 mol.% BIS) is similar to values found for significantly higher crosslinked CS microgels (5 mol.% BIS) in literature.[50] This difference can be attributed to the much larger dimensions of the CS microgels in the present work. We assume that the contact radius $a$ scales approximately with $R^{2/3}$ which is the scaling known from the contact deformation model by Johnson, Kendall and Roberts (JKR-model).

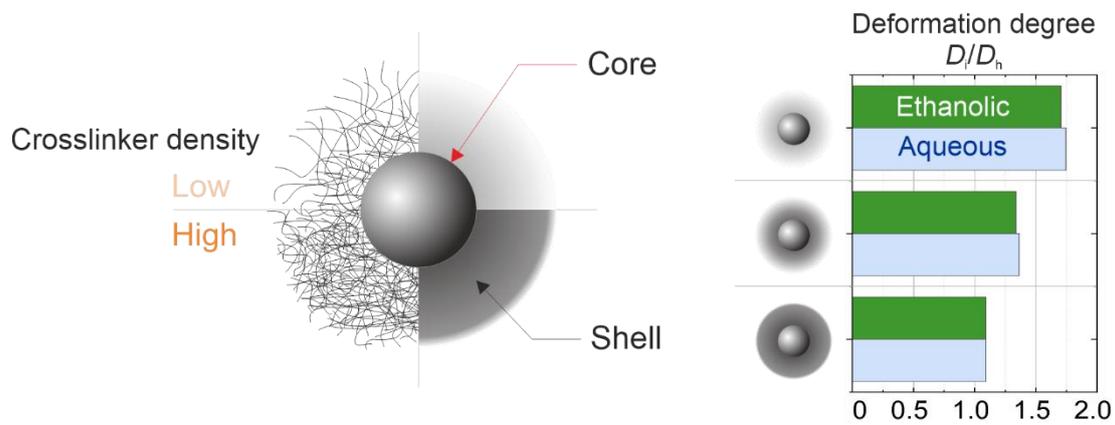

*Figure 1*. CS microgels with shells of different softness. **Left**: *Schematic illustration of the structure of CS microgels with different crosslinker densities. The solid core (silica, illustrated by the grey sphere) is surrounded by crosslinked PNIPAM shells that are simplified by grey coronas. Dark and light coronas denote high and low crosslinking density, respectively.* **Right**: *Degree of lateral deformation at the air/water interface ($D_i/D_h$). The microgels were adsorbed to interfaces from ethanolic and aqueous dispersion, with a 30-minute equilibration. The crosslinker density of the PNIPAM shells increases from top to bottom with the sketches representing the three different CS microgels (CS-low, CS-medium, and CS-high).*

Using ethanol as a spreading agent,[56] the CS microgels were adsorbed at the air/water interface of a Langmuir-Blodgett trough (Microtrough G2, Kibron Inc.) and a surface-modified solid substrate was immersed (90° to the interface and parallel to the barriers) using a dipper.

For the preparation of hydrophilic substrates, standard microscopy glass slides were thoroughly cleaned and subsequently plasma-treated right before the experiment. For the hydrophobic substrates, the glass slides were RCA-cleaned[48] and fluorinated via chemical vapor deposition. Details regarding the microgel synthesis, Langmuir-Blodgett deposition, and substrate preparation are in the **Experimental Section**. The microgel-laden interface was later compressed uniaxially by means of lateral barriers to reach a given Π. Three values of Π are considered to achieve various $D_{c-c}$ (i.e., compression state): approximately 10, 20, and 30 mN/m, to which we will refer as 'low', 'mid', and 'high Π'. The corresponding compression isotherms for all three CS microgel systems can be found in the **Supporting Information** (**Figure S1**). Video microscopy of the drying interface was performed by transferring the microgel-laden interface rapidly (dipper speed: 279 mm/min) onto the solid substrate at constant surface pressure,[53, 57-58]. Images of the drying microgel-laden interfaces (frequently known as microgel monolayer or microgel films) were taken *in situ* immediately after the transfer using reflected brightfield microscopy. Even at a high Π, i.e., high packing fractions, individual CS microgels remain well-resolved due to the thick PNIPAM shells and the silica cores providing large center-to-center distances and high optical contrast. Under our imaging conditions, the drying microgel films also exhibited iridescent colors resulting from thin film interference.[22] These colors can be used to map the local height of the microgel-laden interface with respect to the underlying substrate (details about this method are given in the **Experimental Section**).

## 2.2 Effect of hydrophilicity / hydrophobicity of the substrate

Using a combination of 3D superresolution fluorescence microscopy and dissipative particle dynamics, Hoppe Alvarez et al.[59-60] and Shaulli et al.[61] have demonstrated that poly(*N*-isopropylmethacrylamide) microgels maintain their native spherical shapes on hydrophilic substrates (against an aqueous phase). Conversely, on hydrophobic substrates, which have lower surface energy and prefer to avoid contact with water, the same microgels undergo significant deformation, creating larger contact areas between the microgels and the substrate. This is attributed to the fact that the contact of a hydrophobic surface with amphiphilic polymer chains is energetically more favorable than with water molecules. Recent simulations of tethered homopolymer chains in planar brushes and block copolymer micelles also revealed energetically favorable substitution of surface-water to surface-polymer contacts on hydrophobic surfaces, promoting polymer adsorption.[62]

**Figure 2** depicts the drying of a film containing CS-medium at mid Π on hydrophilic (**Figure 2,** panels **A1-F1**) and hydrophobic (**Figure 2,** panels **A2-F2**) substrates at various stages of the drying process (the corresponding **Videos** are in **SI**). Confocal scans of the meniscus cross-section are used to measure the slope of the meniscus during drying (see **Experimental Section**), i.e., the receding wetting angles $\theta_1 < 1°$ (hydrophilic substrate, **Figure S2**) and $\theta_2 \approx 2°$ (hydrophobic substrate, **Figure S3**). Panel **A1** of **Figure 2** shows the microstructure of the microgel film on the inclined interface, which is indicated by the color gradient. Following this stage, we report the formation of a thin fluid layer where the microgel-laden interface aligns parallel to the solid surfaces, evidenced by the positioning of all microgels in the focal plane (panels **B1** and **B2**). As the level of water lowers further (panels **B**-**D**), the color of the core region (i.e., silica core plus swollen microgel around it) becomes increasingly distinct from the rest, signifying different surface elevations. On hydrophilic substrates, regions with low microgel concentration (empty areas) grow in size as the drying proceeds (hereinafter called hole formation, see **Figure 2**, panels **C1**-**E1**), pushing the microgels against each other until the film completely dries (**Figure 2**, panel **F1**). This hole formation appears to be largely influenced by fluid dynamics, as evidenced by the circular shape of the enlarging holes, which are also observed for both micron-sized hard sphere-like systems[10] and microgel systems[63].

Remarkably, the migration (XY displacement) of the microgels is negligible when the drying takes place on hydrophobic substrates (**Figure 2**, panels **C2**-**F2**). The characterization is repeated at different values of Π. The corresponding microscopy images are depicted in **Figure S4.** We quantify the effect of water evaporation on the microgel films by measuring $D_{c-c}$ before and after the drying (**Figure S5**). $D_{c-c}$ changes significantly when the underlying substrate is hydrophilic indicating higher mobility as compared to the hydrophobic case. This is in general agreement with the findings of Hoppe Alvarez et al.[59-60]

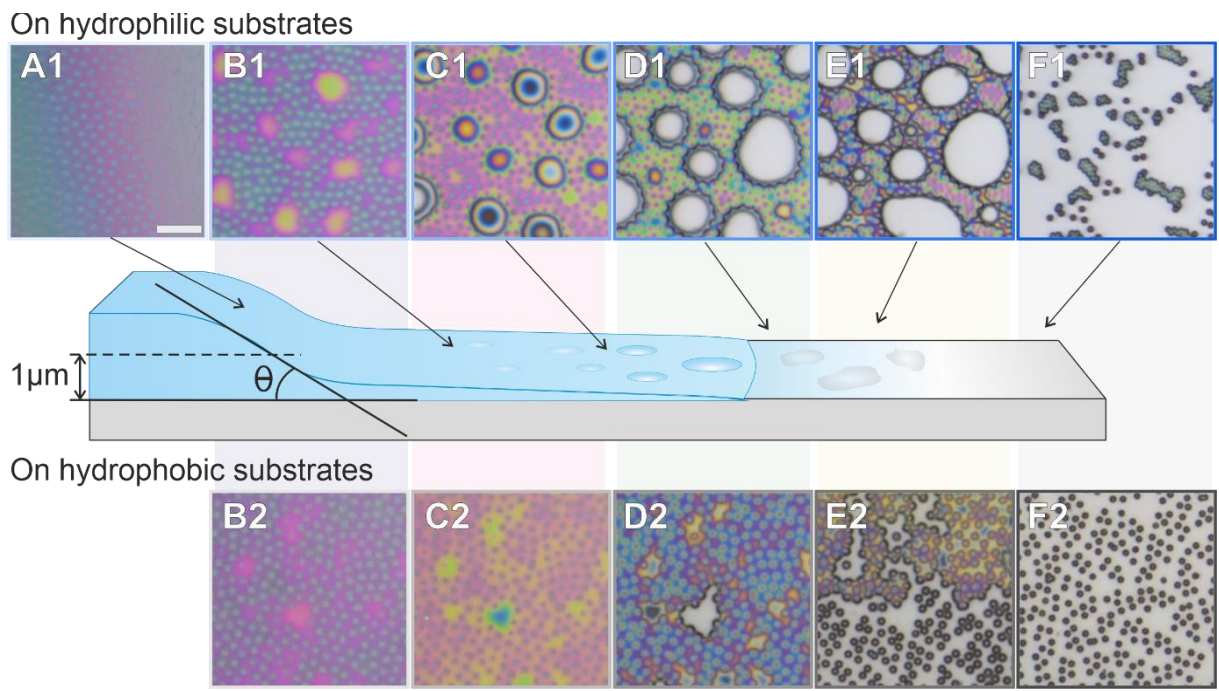

*Figure 2. Sketch illustrating the drying of microgel films (middle row) and reflected light microscopy images of CS-medium films at mid Π (~20 mN/m) transferred onto hydrophilic (A1-F1) and hydrophobic substrates (B2-F2). The colors are due to thin film interference and will be later used to determine the local height. The scale bar corresponds to 5 μm. All microscopy images show the same magnification. θ is the receding wetting angle of the meniscus. Equal letters refer to similar film thicknesses deduced from similar colors.*

To better understand and control the drying and rewetting process, we mimicked the contact scenarios between the microgel-laden interfaces and solid substrates using a modified thin film pressure balance (TFPB) equipped with a porous glass plate and connected to a capillary tube filled with water (**Figure 3A** and **Figure S6**) similar to the one used by Ciunel et al.[64] and Schelero et al.[65] The glass plate was soaked with water and had a truncated hole, where the air/water interface was created on the top and the substrate was placed on the bottom. In analogy to the CLSM measurements, the microgel dispersion was spread at the air/water interface. The porous glass plate was placed in a pressure cell and by varying the air pressure, the setup enabled the on-demand position control of the microgel-laden interface, which extended from a few mm above the solid substrate down to thin wetting films comparable to the microgel size. In the presence of a hydrophilic substrate (**Figure 3B**), i.e., weak microgel-to-substrate adhesion, microgels trapped in the thin fluid film could be released back to the air/water interface on demand by refilling the wetting film with water via a decrease of the pressure in the cell. This phenomenon was most clearly observed in the **Video** that is provided as **Supplementary Material**, where the mobility of the microgel-laden interface is evident as it floats back up. Conversely, the TFPB experiment conducted on hydrophobic substrates (**Figure 3C**) revealed that the microgel monolayer was trapped immediately due to the strong adhesion. Refilling the wetting film with water did not lead to desorption and floating back of the microgels to the air/water surface was not possible. Instead, when attempted (repeatedly increasing and decreasing the position of the microgel-laden interface), it resulted in multilayers of the microgels, see **Video** that is provided as **Supplementary Material** and **Figure S7** in the **Supporting Information.** This spontaneous and favored adsorption of microgels from the water phase onto a hydrophobic substrate was also reported in the drying experiment on microgel-laden droplets.[66]

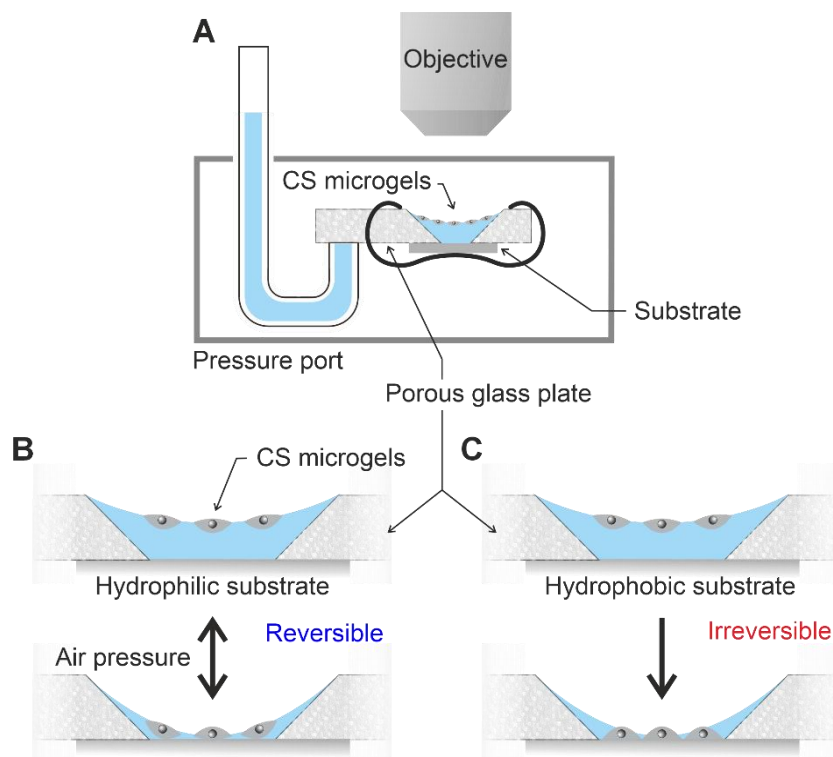

*Figure 3*. Schematic illustrations of a modified thin film pressure balance (TFPB) setup (*A*) and the microgel-laden interface oscillation experiment by pressure modulation using a hydrophilic substrate: far away from the solid substrate (*B*) and in the thin film on the substrate (*C*). Videos recorded for experiments on a film of CS-medium are provided in the *SI*.

Using molecular dynamics simulations with a coarse-grained model and explicit solvent, we further verified the drying scenarios of microgel-laden interfaces (details are presented in **Supporting Information**). Specifically, we constructed a system based on the experimental results in which a monolayer of 16 CS microgels was created (**Figure S8**). **Figure 4** illustrates the microgel deformation at the air/water interface before water evaporation (panels **A, A1**) and its shape on hydrophilic (panels **B, B1**) and hydrophobic (panels **C, C1**) substrates after water evaporation.  The term "water" in our description refers to a simplified model, specifically a Lennard-Jones truncated and shifted (LJTS) fluid. **Figure 4** presents two significant aspects: (1) the lateral size of microgels decreases upon drying, and (2) the contact area of dry microgels on hydrophobic substrates is larger compared to hydrophilic substrates.

The enhanced contact area on hydrophobic substrates persists at all stages of the microgel adsorption, from the initial solvent-rich stage to after drying, resulting in stronger molecular "friction" between the microgel and substrate. These findings support the experimentally observed structural quenching (maze-like cracks) upon solvent evaporation and the reduced mobility of microgels on hydrophobic substrates.

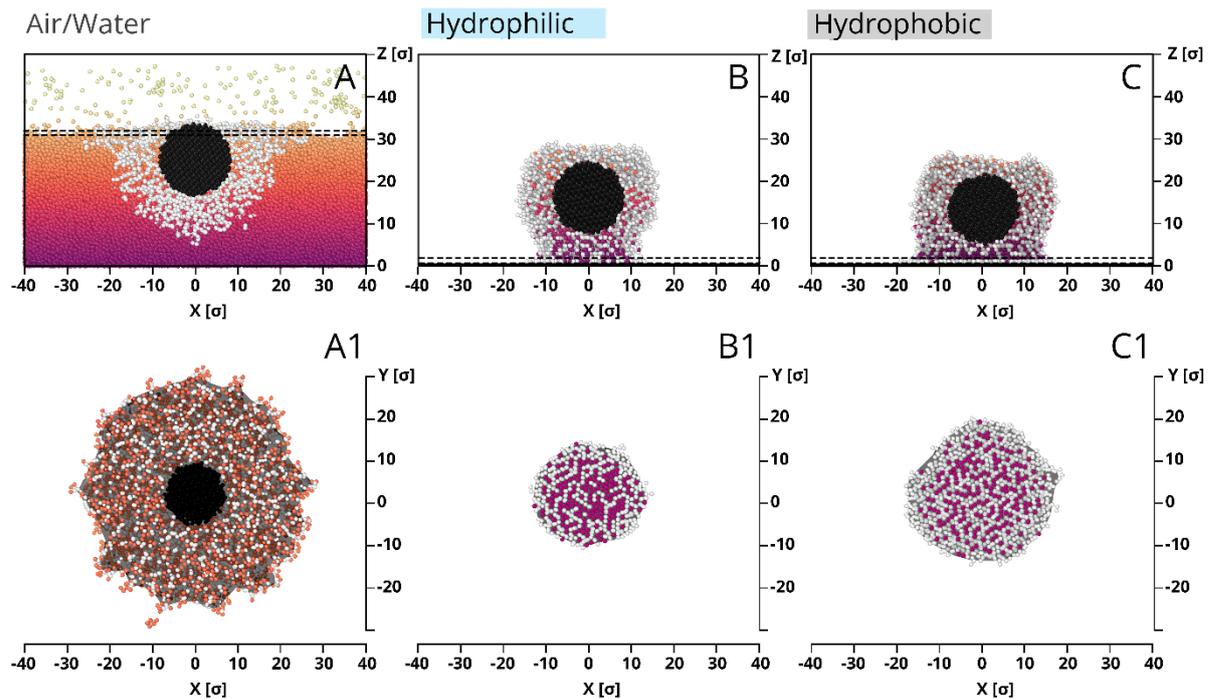

*Figure 4. Snapshots of a selected CS microgel in the monolayer obtained at low Π before (**A,A1**) and after drying on hydrophilic (**B,B1**) and hydrophobic (**C,C1**) substrates. The upper row demonstrates a side view (cross-section through the center of mass). The bottom row depicts thin layers of the microgels shown by dashed lines in the upper row, i.e., they are upper view (**A1**) and contact area on hydrophilic (**B1**) and hydrophobic (**C1**) substrates. The purple dots in (**B,B1**) and (**C,C1**) correspond to residual water.*

Analogous to **Figure 2**, the simulation-assisted drying behaviors of the microgels on differently modified substrates (hydrophobic vs. hydrophilic) during a continuous water evaporation process are illustrated in **Figure 5** (panels **A**-**E**). It is important to note that the evaporation of water from the interface is not uniform, with the most rapid drying occurring in the regions between the microgels, where a lower polymer concentration enables efficient evaporation. This asymmetry results in the formation of a drying front (hole formation), prompting XY displacement of microgels. The hole formation could be further influenced by disjoining pressure, which stabilizes the thin water film on the substrate. On hydrophilic substrates, the disjoining pressure is positive, which stabilizes the film and prevents local film instabilities. However, for hydrophobic substrates, the disjoining pressure should become negative, as previously reported,[67] resulting in an unstable film. In other words, at a critical thickness of the film (generally thicker on the hydrophobic substrate), the evaporation of water triggers the instability leading to the de-wetting of the water film.

The simulation results show that the mobility of the microgels on the hydrophobic substrate is reduced (**Figure 5**, panels **A2**-**E2**). Here, we also see that the microgel cluster in a dry state (**Figure 5**, panel **E2**) is less compact compared to the case on the hydrophilic substrate (panel **E1**). The compactization, which is driven by the minimization of the surface energy of the drying microgels, proceeds faster on the hydrophilic substrate due to the higher mobility of the microgels. The discrepancy in microgel behavior on various substrates is quantified by the parameter $D_{c-c}$ in **Table S4** in the **SI**. However, it is evident that the microgel-to-substrate adhesion is insufficient to prevent XY displacement in both cases. This might be because water evaporation in the experiments leads to a far out-of-equilibrium quenched (or "frozen-in") state of the system, whereas in the simulations, interaction parameters used allow equilibration of the system even in a dry state. See **Model and Simulation Details** in **SI** for more details (**Figure S9-14**).

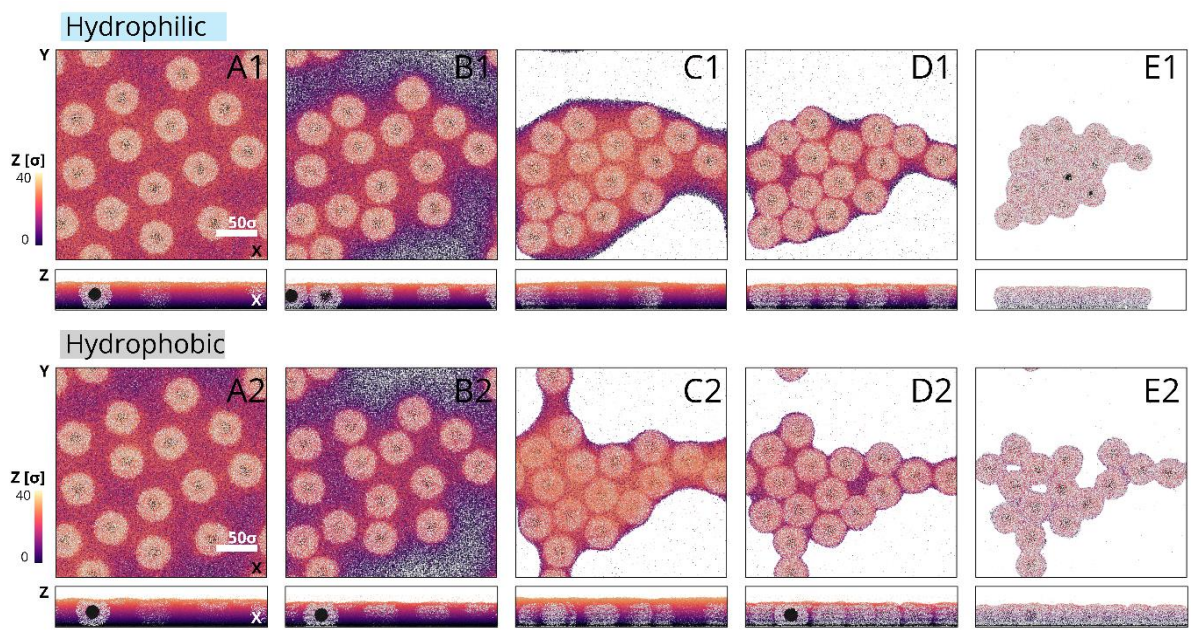

*Figure 5*. Illustration (top views) of the different stages of microgel film obtained at low Π drying on hydrophilic (**A1-E1**) and hydrophobic (**A2-E2**) substrates as revealed by computer simulations. The narrow panels below the top views are the corresponding side views, i.e., cross-sections of each panel along Z. The height of the film with residual water is depicted by the vertical color bar. White regions in the snapshots correspond to the bare substrate after water evaporation.

## 2.3 Effect of microgel softness

We take a further step to characterize the drying of soft films and compare two CS microgels with different crosslinker densities, i.e., softness. **Figure 6** illustrates different stages of the drying process of CS-low and CS-high microgel films at mid Π. For "softer" and thus more deformable microgels (CS-low, **Figure 6,** panels **A-D**), the surface chemistry of the substrate has minimal-to-no influence on the spatial rearrangement upon drying. Significant microstructural changes, as reported for CS-medium films (**A1**-**F1** of **Figure 2**), occur for CS-low only at high Π on hydrophilic surfaces (**Figure S15**). Conversely, microgels with "harder", less deformable shells (CS-high, **Figure 6E-H**) undergo significant XY displacements as they dry, irrespective of the hydrophobicity of the substrate. In particular, an abrupt and sudden collapse of microgels (i.e., a fast and large hole formation) is observed when the film dries onto hydrophilic substrates and the thin layer of fluid reaches a certain thickness. On the other hand, on hydrophobic substrates, the microgels tend to lean and/or slide towards one another, creating maze-like cracks. In analogy to **Figure S4** and **S5**, we also report in **Figure S15-17** the microscopic images and $D_{c-c}$ for different Π values. (In the presence of clusters $D_{c-c}$ refers to the distance between the microgels in the clusters.)

To elucidate whether the observed effects rely on the core-shell morphology of our microgels, which affects at least the total deformability of the microgels, we also studied purely polymeric microgels of similar and smaller total sizes. One would anticipate that the latter microgels would deform more easily and thus exhibit lower mobility during drying. However, similar changes in microstructure as for our CS microgels during drying were observed for pure PNIPAM microgels of comparable sizes (**Figure S18**, **S19**) as well as smaller CS microgels (**Figure S20**) in the mid-high Π regime.

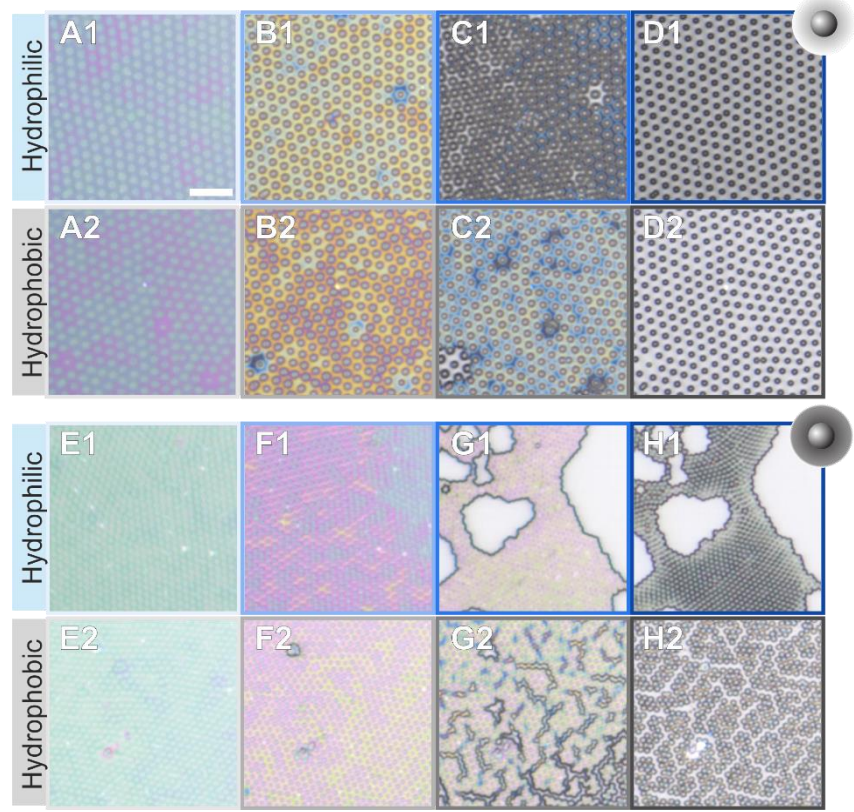

*Figure 6. Reflected light microscopy images of CS-low (**A-D**) and CS-high microgel films (**E-H**) at mid Π (~ 20 mN/m) transferred onto hydrophilic (**A1-D1, E1-H1**) and hydrophobic substrates (**A2-D2, E2-H2**) during the drying process. The scale bar corresponds to 5μm.*

### 2.4 Interplay between the adhesion and immersion capillary force

**Qualitative picture.** When a solid particle is immersed in the liquid film, it is under the influence of many forces, such as gravity, and forces that come from the presence of the liquid around the particle, such as buoyancy and surface tension. For small particles, e.g., diameter below 1 μm (which is the case we are investigating), the influences of gravity and buoyancy become negligible. Thus, only the force arising from the surface tension, $F_\sigma$, of the liquid along the wetted perimeter (and the upward normal force exerted from the solid substrate, $F_N$, becomes relevant, where their force balance is given by:[68]

$$F_N = 2\pi R\sigma \sin\phi_c \sin(\phi_c + \alpha)$$

where $R$ is the radius of the particle, $\sigma$ is the surface tension of the liquid, and all contact angles involved, governed by the wetting properties of the particle surface, namely the meniscus slope angle, $\psi_c$ not to be confused with the receding wetting angle θ from **Figure 2**, particle contact angle, α, and the wetting angle $\phi_c$ where $\psi_c + \alpha + \phi_c = 180$,[68] as depicted in **Figure 7**. In the presence of another particle (same type and size) in the vicinity, the particles are under the influence of the immersion capillary force. The lateral capillary interaction between the two particles can be described as:

$$\Delta W \approx -2\pi\sigma Q^2 K_0(qD_{c-c})$$

where the "capillary charge" is $Q \equiv R_c \sin\psi_c$ and $K_0$ is the modified Bessel function of the second kind and zeroth order, in addition:

$$q^2 = (\Delta\rho g - P_d')/\sigma$$

Where $\Delta\rho$ is the difference between the densities of the two fluids (air and water) and $P_d'$ is the derivative of the disjoining pressure as a function of the height of the film.[10-11] The disjoining pressure becomes negligible when $H_0$ exceeds 100 nm for the wetting film.[9] The immersion capillary force exerted on the two particles, which is the derivative of the lateral capillary interaction:[10]

$$F = -\frac{d\Delta W}{dD_{c-c}} \approx -2\pi\sigma Q^2 q K_1(qD_{c-c})$$

where $K_1$ is the modified Bessel function of the first order.

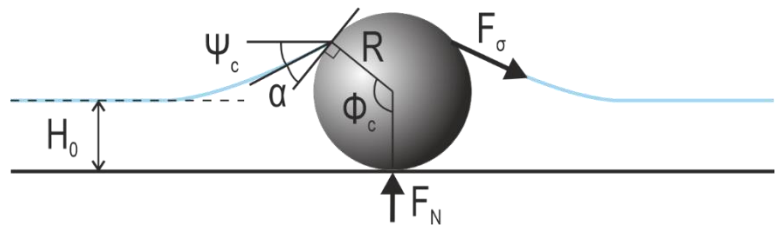

*Figure 7. Scheme of forces acting on a small particle with a radius of R, immersed in a thin water film with a height of $H_0$. The angles between the interfaces can be characterized by the particle contact angle, $\alpha$, and the wetting angle $\phi_c$. The particle is under the influence of forces that arise from the surface tension, $F_\sigma$, of the liquid along the wetted perimeter and the upward normal force exerted from the solid substrate, $F_N$.*

For rigid spheres with dimensions relevant to our CS microgels, i.e. $R_c < R < R_h$, the lateral capillary interaction ($-\Delta W$) reaches theoretical values that exceed the thermal energy by up to eight orders of magnitude, depending on $\psi_c$ and $D_{c-c}$. Since our CS microgels consist of a solid core surrounded by a soft and deformable microgel network where the contact and meniscus slope angle cannot be easily determined and potentially also depend on the water content in the microgel shell, capillary interactions are expected to significantly differ from that of rigid spheres. In a recent study on the drying of microgel dispersions within capillary tubes a drying behavior in between of colloidal dispersions and polymer solutions has been found.[69] Due to the microgel's abilitiy to deform, deswell and interpentrate, their drying behavior from bulk dispersion is dominated by molecular scales. To some extent we expect this to be also relevant for drying of monolayers confined in a thin liquid film.

To elucidate the physical phenomena that occur during the drying of our microgel films, we first illustrate how we picture the microgel structure and deformation at air/water interfaces for "softer" and "stiffer" CS microgels, as depicted in **Figure 8** panels **A1** and **B1**, respectively. Our understanding described here will be quantified and further validated in the following section. For simplicity, the microgels can be considered as made of three parts: I-part (which interacts with the air/water interface), B-part (which interacts with the bulk water phase), and S-part (which interacts with the solid surface). Due to the lateral deformation (see e.g., **Figure 1**) caused by interfacial tension, leading to non-isotropic softness within the polymeric network, the portion at the interface (I-part, **Figure 8** panels **A1**, **B1**) experiences more stretching than the portion in bulk (B-part).[54]

As water evaporates from the thin fluid layer (see e.g., **Figure 2**, panels **C-D**), the bottom B-part of the shells comes in contact with the underlying substrate (S-part, **Figure 8** panels **A2**, **B2**). If the resulting adhesion (dominated by the amount of material interacting with the substrate) is not strong enough to counteract the film instability or the de-wetting of water, microgels migrate (hole formation) and aggregate (see e.g., **Figure 2**, panels **D1, H1**). If the adhesion is strong enough, microgels stay in position in the thin aqueous film, where the local meniscus slope angle, $\psi$, and microgel-to-substrate contact area (colored orange) increases while the average film thickness decreases as the drying proceeds, increasing the immersion capillary force exerted among microgels. At the same $D_{c-c}$ and film height ($D_h > H > D_c$), the harder microgels will be under a stronger immersion capillary force due to the steeper meniscus slope angle ($\psi_{harder} > \psi_{softer}$) and have weaker microgel-to-substrate adhesion (i.e., smaller contact area) due to the high elastic modulus of the polymer network. At a certain height, the immersion capillary force (long-ranged) and disjoining pressure (short-range) of the film can surpass the microgel-to-substrate adhesion, causing the film to rupture or de-wet, i.e., microgels to migrate (distance traveled > $D_{c-c}$) or collapse onto one another (distance traveled < $D_{c-c}$, **Figure 8** panel **B3**), i.e., there is a gradual decrease in $D_{c-c}$ as the effective volume of microgel reduces due to evaporating water. If the adhesion persists stronger than the acting immersion capillary force, the migration of microgels does not occur (**Figure 8** panel **A3**, no XY displacement) and the microstructure of the microgel-laden interface is better preserved. Therefore, if the goal is to produce dried replicas of the interfacial microstructure, soft microgels, and hydrophobic (or oppositely charged) substrates are to be used rather than harder microgels and hydrophilic substrates under fast drying conditions.

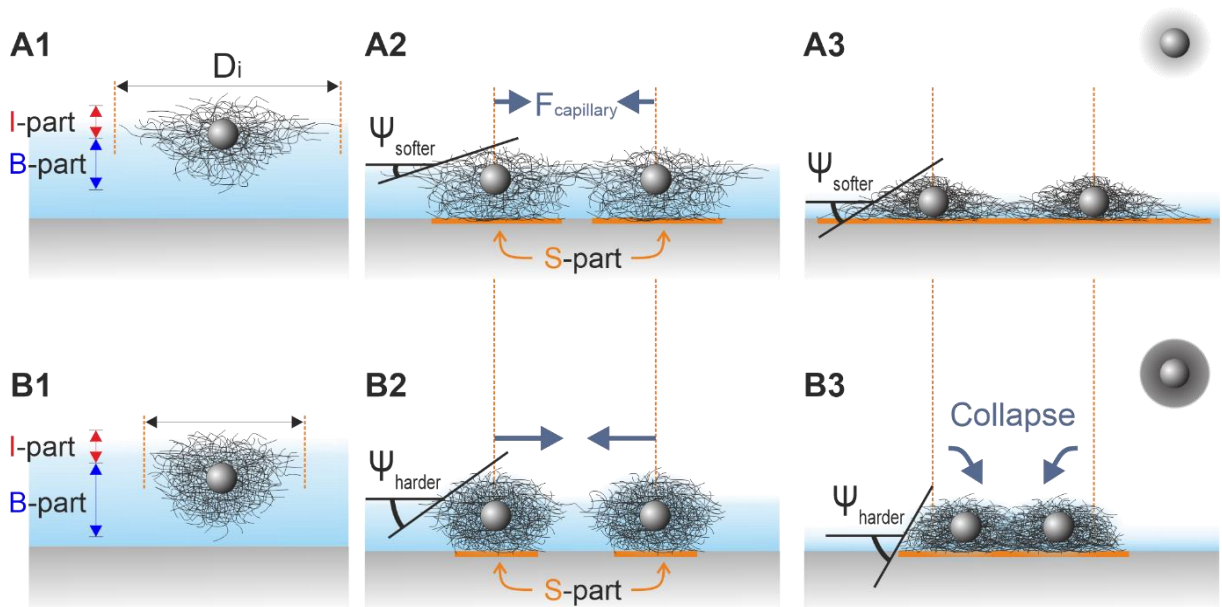

*Figure 8. (A1-B1) Schematic illustration of microgels with a low (A1) and high crosslinker density (B1) at the air/water interface and their interfacial diameter, $D_i$. (A2-B2) Upon drying, at the same $D_{c-c}$ and a given film height ($D_h > H > D_c$), softer microgels (A2) have a larger contact area to the substrate (highlighted in orange) and lower meniscus slope angle (ψ) compared to harder microgels (B2). (A3-B3) Soft microgels are, therefore, more likely to stay in position (A3) than to collapse (B3).*

**Quantitative validation: Thin film interference and particle tracking.** For a more quantitative validation, we measure film height, *H*, during the drying of the microgels after the formation of a thin fluid layer. We consider interfaces filled with CS microgels of different softness (CS-low, CS-medium, and CS-high) at a comparable $D_{c-c}$ (≈ 840 nm, at the air/water interface), and extract the instantaneous velocity *V* of the particles (i.e., the displacement of the core's center between consecutive frames) and the height of the microgel film as it dries. The height profiles of the drying films were traced by their apparent colors, which stem from thin film interference,[22, 70] under the assumption that the interface is perfectly flat, and the refractive index of the thin film equals that of water throughout the drying process. However, it is important to note that as the drying progresses, the height profiles will be increasingly underestimated due to the increasing effective refractive index of the film (increasing polymer volume fraction). The high refractive index and large size of the silica cores introduce further complexity. To simplify the calculation, we masked the core areas (arbitrarily determined, approximately 600 nm in diameter) and only considered the height of the shell area (*H* in **Figure 9A**) for radial averaging. We define $H = H^*$ (critical height), when the microgels start to move from their original XY position. The methodology is described in detail in the **Experimental Section**.

**Figure 9** (panels **B**-**D**) shows the evolution of the height profiles of microgel-laden interfaces during drying on hydrophilic (blue) and hydrophobic (grey) substrates as a function of a normalized time, where *t* = 0 is the onset of XY displacement of the individual microgels. The corresponding instantaneous velocities are in **Figure S21**. All height profiles exhibit plateau regions near $H = D_c$, possibly due to the combination of (1) the elasticity of the microgel networks supported by the core, (2) increased polymer volume fraction, (3) artifact of radial averaging (due to the changes in ψ as the drying proceeds). Notably, the height of the microgel films is generally lower on hydrophobic substrates compared to hydrophilic substrates because of the stronger microgel-to-substrate adhesion and more pronounced deformations of the polymer networks in contact with the solid surface. The critical height $H^*$ is also consistently higher for interfaces made of microgels with "harder" shells, as illustrated in **Figure 9** (panels **B**-**D**) and **Table S1**. The "harder" polymer network appears to resist the thinning of the film, which could also result in a relatively smaller contact area with the substrate and, consequently, weaker microgel-to-substrate adhesion and higher mobility during drying. The difference in softness among CS microgels is also evident in the lateral deformation rate at the air/water interface in **Figure 1**, **Figure S1**, and **Table S1**. Our results indicate that the drying of a microgel film involves an intricate interplay among various factors, such as microgel-to-substrate adhesion, microgel compression degree, immersion capillary forces, and others e.g., free energy associated with interface formation, Marangoni flow, drying conditions.[71-72] Further quantification of the phenomena requires values of the slope angle of the meniscus and microgel adhesion both of which will change as the drying processes.

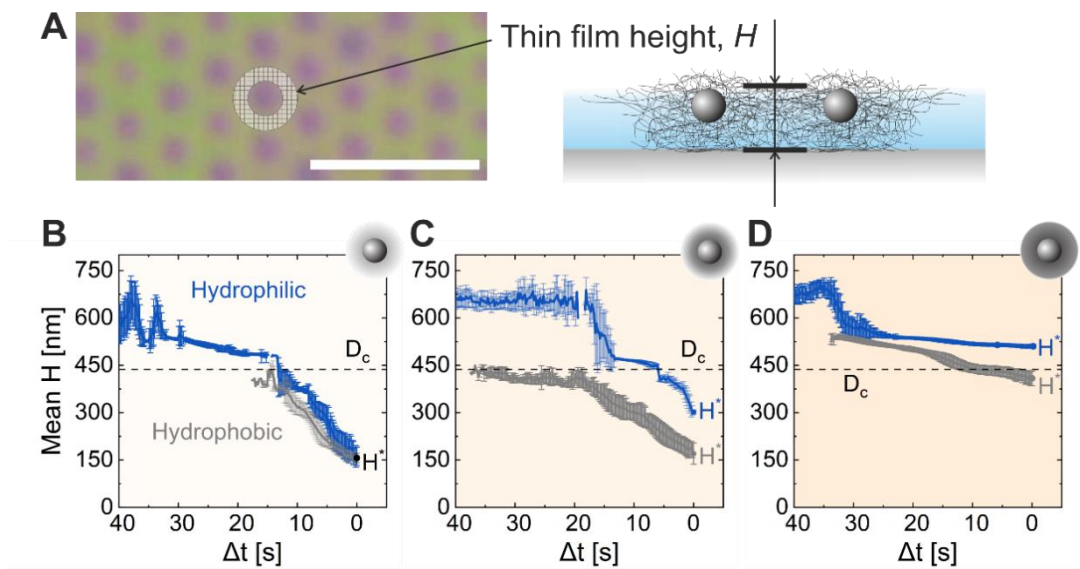

*Figure 9. Film height, H, shown in an exemplary frame (A, left) and in a sketch (A, right). The scale bar corresponds to 5 μm. Evolution of the height profiles of ten randomly chosen CS microgels in corresponding films drying on the hydrophilic (blue) and hydrophobic (grey)*

*substrates for CS-low (**B**), CS-medium (**C**), and CS-high (**D**). Δt represents the time difference, with 0 as the reference point, where H\* is observed.*

## 3. Conclusion

Interparticle interactions among soft micro-units and their phase behavior at interfaces are of great importance in fundamental interface science, biophysics, and additive manufacturing. Studying such phenomena at fluid interfaces, however, is challenging due to the difficulty in visualizing both the building blocks and the deformed interface, particularly in condensed (or highly compressed) states. Our results provide direct imaging and quantification at the single-particle level of drying model soft interfaces consisting of microgel monolayers on various solid substrates. Using variously crosslinked CS microgels, we demonstrated that the resulting dried microstructure of the microgel monolayers is significantly influenced by the softness of the microgels and the wettability of the substrates. In general, microgels with higher softness better maintain the original assembled structure due to more pronounced deformation at the interfaces. This results in lower capillary forces due to a less curved air/water interface. In addition, increased hydrophobicity of the substrate supports the conservation of the arrangement of the microgels during drying due to enhanced adhesion to the substrate and, consequently, lower mobility during the drying process.

We experimentally showed that the drying dynamics of such monolayers involve an intricate interplay among various factors, including microgel-to-substrate adhesion, immersion capillary forces, and the free energy associated with the interface formation. Our results are supported by molecular dynamics simulations, where the hole formation and the varying degrees of deformation, causing differences in their drying dynamics on differently surface-modified substrates upon solvent evaporation, are visualized. This implies that the interpretation of the "2D" assembly of soft colloids at interfaces must consider the often-overlooked 3D aspects:

how microgels deform in 3D initially at interfaces, undergo further deformation under compression, and continue to deform during the drying process. This perspective has also been emphasized in a recent study on emulsion stabilization using microgels.[73] While the presence of the rigid, non-deformable core increases the operative immersion capillary force and may influence the shell's effective softness, when trapped in thin liquid films (i.e., during drying on a substrate), the deformation of the interface is inevitable, even for submicron-sized purely polymeric microgels without rigid cores. The system will then react on the microgels to minimize the surface energy configuration.[74] Therefore, *ex-situ* results, i.e., data obtained after transfer onto a substrate, should be approached with caution.

To date, there is no generally accepted model for the interactions among soft colloids at interfaces bridging from dilute to condensed (compressed) state. In future investigations, we will explore the interparticle interactions of these soft colloids, especially at the single colloid level, using techniques such as optical tweezing. This approach has the potential to enhance our understanding of the complex phase and rheological behaviors exhibited by soft colloids. Furthermore, for applications where soft materials come in contact with more rigid surfaces and undergo dehydration and rehydration (e.g., tissue and food engineering, biomedical application),[75-78] it could be interesting to explore the influence of varying surface roughness or porosity and measure the effective lateral attraction force among differently engineered microgels and their microgel-to-substrate adhesion forces in various environment.

## 4. Experimental Section

### 4.1 Materials

Ethanol (Sigma-Aldrich, 99.8%), ethanol (Heinrich-Heine-University, chemical store, p.a.), tetraethyl orthosilicate (Sigma-Aldrich, 98%), ammonium hydroxide solution ($NH_3$ (aq.), PanReac AppliChem, 30%), ammonium hydroxide solution ($NH_3$ (aq.), VWR, 25%), ammonium hydroxide solution (NH$_3$ (aq.), PanReac AppliChem, 30%), hydrogen peroxide solution (H$_2$O$_2$, Fisher Chemical, 30 wt %), rhodamine B isothiocyanate (Sigma-Aldrich, mixed isomers), methacryloxyethyl thiocarbamoyl rhodamine B (MRB, Polysciences, Inc.), (3-aminopropyl)trimethoxysilane (Sigma-Aldrich, 97%), 3-(trimethoxysilyl)propyl methacrylate (MPS, Sigma-Aldrich, 98%), 1H,1H,2H,2H-perfluoroctyltriethoxysilan (PFOES, J&K Scientific, 97%), *N,N'*-methylenebisacrylamide (BIS, Sigma-Aldrich, 98%), and potassium peroxodisulfate (KPS, Sigma-Aldrich, 99%) were used as received. Water was purified by a Milli-Q system (18.2 MΩ cm) and *N*-isopropylacrylamide (NIPAM, TCI, 97%) by recrystallization from cyclohexane (Fisher Scientific, 99.8%).

**4.2 Synthesis**

The synthesis of CS microgels was done via seeded precipitation polymerization. In brief, we synthesized silica particles using the well-known Stöber procedure and modified their surfaces with MPS. RITC dye was incorporated into the particles. The diameter, $D_c$, measured by transmission electron microscopy (TEM) was 437±20 nm. The PNIPAM shell encapsulation of the cores proceeded through seeded precipitation polymerization at 70 °C with various feed concentrations of BIS relative to NIPAM. Specifically, 1.0, 2.5, and 7.5 mol.% BIS were used for the synthesis of CS-low, CS-medium, and CS-high, respectively. The exact amounts of chemicals used are listed in **Table 1** and the synthesis protocols for the micron-sized CS microgels are detailed elsewhere.[48] After purification via repeated centrifugations and re-dispersion cycles, the dispersion was freeze-dried, re-dispersed in ethanol at a concentration of 5 w/v.%, and left overnight on a 3D shaker before use.

**Table 1.** Respective amounts of chemicals used for the synthesis of CS-low, CS-medium, and CS-high.

| Sample name | Core | | Core-shell | | | | |
|---|---|---|---|---|---|---|---|
| | Particle /ethanol [g/ml] | added [ml] | $H_2O$ [ml] | NIPAM [g] | BIS [g] | KPS [g] | $D_h$@ 20°C [nm] |
| CS-low | 0.53 | 1 | 255 | 1.0014 | 0.0135 | 0.0256 | 1114.1 |
| CS-medium | | | | 1.0012 | 0.0343 | 0.0255 | 1003 |
| CS-high | | | | 1.0004 | 0.1022 | 0.0258 | 1053.2 |

The pure PNIPAM microgels were synthesized via precipitation polymerization according to previously published work.[19-20] The synthesis protocols were modified for dye incorporation for smaller microgels (*s*MG, $D_h \approx 800$ nm) (**Figure S4**) as described in [42] as well as for larger microgels (*l*MG, $D_h \approx 1,500$ nm) (**Figure S5**). *l*MG was synthesized with 2.11 g of PNIPAM and 60 mg of BIS (2 mol.%), dissolved in 125 ml of water. The mixture was injected through a 0.2 µm Nylon syringe filter into a three-neck round-bottom flask equipped with a reflux condenser and a magnetic stirrer. 250 µL of MRB dye aq. solution (1 mg/ml) was added to the flask. The mixture was heated to 45 °C and equilibrated for an hour while purged with nitrogen under stirring. 0.1054g KPS in 5 ml water was added to the flask through a 0.2 µm Nylon syringe filter. After initiation, the temperature was ramped up to 65 °C in 40 minutes. The reaction was kept overnight under stirring at 65 °C. The dispersion was then filtered through glass wool. The synthesized PNIPAM microgels were dialyzed against water for two weeks, freeze-dried, and re-dispersed in ethanol (1 w/v.%).

### 4.3 Methods

**Dynamic light scattering (DLS).** The hydrodynamic diameter, $D_h$, of the CS microgels was determined using a Zetasizer Nano S (Malvern Panalytical). The device is equipped with a HeNe laser (4 mW, 633 nm) and a temperature-controlled jacket for the sample. Measurements were performed at a scattering angle of 173 ° at 20 °C. Three measurements were performed.

Values of $D_h$ reported are averaged from the z-averages obtained from the measurement software.

**Glass substrates for monolayer transfer.** For the preparation of glass substrates with a hydrophilic surface, standard microscopy glass slides were thoroughly cleaned and rinsed using water and ethanol, and then plasma-treated prior to the monolayer transfer. For the hydrophobic surface modification, the glass substrates were RCA-cleaned[79-80] and surface-modified via chemical vapor deposition. 200 µl of PFOES was stored with the cleaned glass substrates in a desiccator overnight under vacuum (25-30 mbar). The glass substrates were then placed in an oven at 120°C for an hour to ensure covalent bonding. Afterward, the substrates were washed in ethanol in an ultrasonic bath to remove unreacted silane molecules. The contact angle with 5 µl water droplet was measured by a drop shape analyzer (DSA 25, Krüss) at room temperature (25.5-26.6 °C, relative humidity 35-46 %). The contact angle of the freshly plasma-treated glass substrate (within 10 mins) was $3.5 \pm 0.8°$ and that of the PFOES-modified glass substrate was $103 \pm 0.9°$.

**Monolayer transfer and optical microscopy.** The transfer of the CS microgel monolayer on the differently modified substrates was carried out using a Langmuir-Blodgett deposition trough (Microtrough G2, Kibron Inc.) equipped with a film balance, two Delrin barriers, a dip coater, and an acrylic cover box. The monolayer deposition was done rapidly maintaining the measured surface pressure (surface pressure changed during the transfer: $1.4 \pm 1.1$ mN/m, compression speed: 187 mm/min, dipper speed: 279 mm/min) positioned at 90 ° to the air/water interface and parallel to the barriers. The "wet" monolayer on the glass slide was then placed under a light microscope (Eclipse LV150N, Nikon) equipped with a camera (DS-Fi3) and a 100× objective for the *in situ* monitoring of the drying of the monolayer (approximately 5 frames per second). Pure PNIPAM microgels were investigated under a fluorescence microscope (Olympus IX73) equipped with a mercury lamp, a fluorescence filter set, a CMOS camera, and a 60× objective.

**Confocal laser scanning microscopy.** XZ *in situ* time series scans of the drying CS microgel monolayers were acquired using a Zeiss inverted LSM880 Airyscan microscope system (Zeiss Microscopy GmbH), equipped with a Plan-Apochromat 40x/0.95 dry objective lens. Microgel monolayers were prepared as described at the air/water interface in a crystallizing dish and transferred on cover glasses as described in [48] and were mounted immediately on the microscope motorized stage. The time series scans were started in regions that were not in the dried state. Two simultaneous acquired channels were set up in fluorescence and reflective mode to observe the Rhodamine B labeled silica cores in the monolayer and their distance to the cover glass and the water meniscus, respectively. 561 nm was used at 3 % intensity as an excitation laser line with a PMT detector set at a range of 580-670 nm for the acquisition of the fluorescence. For simultaneous acquisition of the laser reflection, another GaAsP detector was set at a range of 540-580 nm overlapping with the 561 nm excitation laser line. The general acquisition parameters were set as the following: The calculated pinhole size was used at 0.44 airy units. The x-axis pixel size was set to 208 nm at a total scan length of 213 µm. The z-axis covered a range of 49 µm as 100 slices with an interval of 492 nm. The scans were performed in line scan fast Z mode at a scan speed of 2.05 µsec/pixel resulting in an average framerate of 1.08 s/frame. The receding wetting angles, θ, of the drying microgel thin films (**Figure S2, S3**) are measured using ImageJ (Angle tool, version 1.53k, National Institutes of Health, USA).

**Thin film pressure balance.** To investigate the adsorption and desorption of microgel monolayers at the water/substrate interface, a custom-built thin film pressure balance (TFPB) using the porous glass plate method[81-82] in wetting configuration[64] was used at 22 °C. The centerpiece was a porous glass plate (pore size 10-16 µm, porosity P16 (ISO 4793)) with a drilled hole of diameter of about 1 mm. The porous glass plate was attached to a glass capillary tube (film holder), which was filled with water in a way that the water could flow from the porous glass plate to the capillary tube and vice versa. The substrate of interest (various hydrophilic glasses, silicon wafers, and hydrophobic glass) was placed underneath the hole of the porous glass plate and was fixed with a stainless-steel clamp, as shown in **Figure S6A**. The monolayers of microgles were prepared as for the confocal microscopy using the whole film holder with the clamped substrate submerged before injecting certain volumes of an ethanolic dispersion of the CS microgels at the air/water interface, which corresponds to a surface pressure of approx. 20 mN/m. The film holder was placed in a sealed stainless-steel pressure chamber with a quartz glass window for the imaging, as shown in **Figure S6B**. The area of interest was illuminated by a cold-filtered halogen lamp through the reflective light microscope optics and imaged by a color CMOS camera (JAI Go-2400-USB, pixel size: 5.86 µm × 5.86 µm, Stemmer Imaging Puchheim). In combination with the optics (reflected light microscope, extension tube), the resolution of the camera system was 1.72 pixel/µm.

**Film thickness during drying.** The drying films with the microgel monolayers started to appear colored due to interference effects when the total film thickness lowers below 1 µm. A model color spectrum was simulated with an algorithm based on a water slab covering a reflective surface (modified from the free-standing water slab color simulation in [22, 70]). The spectrum was represented in the hue, saturation, and value (HSV) color space, and is stored in thickness steps of 1 nm in a lookup table. Each film thickness in the range from 100 to 1000 nm has a corresponding set of unambiguous HSV values. Pixel by pixel the hue values of the film image were automatically compared with the lookup table and the corresponding film thickness for each pixel is the result. The height of their shell's shoulder ($H$, **Figure 9**) was traced as the thin fluid layer dries. For each image and each particle, $H$ is calculated using the average color of a circular region around the core, stemming from thin film interference (note that the areas occupied by the cores themselves are excluded from the analysis).

**Particle tracking.** The detection of the onset of the XY displacement was achieved via particle tracking of the individual microgels in terms of velocity. The averaged velocity over all microgels in the frames does not allow for the precise detection of the onset of the XY displacement because the length and height of the microgel thin film varied in all cases, leading to many frames containing both swollen and collapsed CS microgels in varying ratios. Consequently, ten CS microgels were randomly selected for analysis.

**Molecular Dynamic (MD) simulations.** The MD simulations were carried out in the canonical (NVT) ensemble using the LAMMPS package.[83] In rectangular simulation boxes, 16 identical core-shell microgels were created near the planar liquid/vapor interface. The monolayers at low and high compression states were prepared and studied. Periodic boundary conditions were applied in xy directions. The system consisted of three types of coarse-grained beads: liquid/vapor beads (W), solid microgel core beads (C), and polymeric beads (S) forming the microgel shell (**Figure S8**, **Table S2**). All beads are Lennard-Jones (LJ) particles having the same characteristic size parameter, σ and mass, m. A solid flat wall (sub), referred to as the substrate, was placed at the bottom of the simulation box (z=0), while a virtual reflective wall was introduced at the top (z=Lz).

The Lennard-Jones truncated and shifted (LJTS) fluid model described via 12-6 LJTS potential was used to simulate the water liquid and vapor phases (details are presented in **Supporting Information**).[84] This highly computationally efficient model accurately describes simple fluids, enabling us to introduce and study the liquid/vapor interface. The vapor-liquid interfacial properties of the model have been extensively investigated in the literature.[85-88] In our model, $\epsilon_{W-W} = 1\varepsilon$, is applied at $T = 0.72\ \epsilon/k_B$.

The substrate was modeled via 12-6 LJTS potential as a smooth, solid, flat wall at the bottom of the simulation box (details are presented in **Supporting Information**). Such "simplification" of the substrate description is reasonable, keeping in mind that the roughness scale of the clean glass surface (hydrophilic) as well as of PFOES coating (hydrophobic) is negligibly small compared to the size of the microgel. To capture the specific hydrophobicity or hydrophilicity of the substrate, the intrinsic contact angle $\theta$ of a water droplet was estimated by variation in the interaction strength parameter $\epsilon_{W-sub}$. It was found that $\epsilon_{W-sub} = 1.7\,\varepsilon$ and $3.8\,\varepsilon$ accurately reproduce intrinsic contact angles of water droplets of approximately 100° and 3°, representing hydrophobic and hydrophilic substrates, respectively.

The core-shell structure of the microgels was designed using the methodology described in our previous works[89] and detailed in the **Supporting Information**. In analogy with the experiment, we designed the microgels with a more crosslinked inner part near the solid core and a fuzzy outer shell. The average fraction of crosslinkers in the polymeric shells equals 9.4% (see Table S3). The polymeric beads (S) – liquid interaction was estimated based on the diagram of states of the droplet, which contains polymer, as obtained in.[90] The combination of parameters $\epsilon_{S-S}, \epsilon_{W-S}$ with a fixed $\epsilon_{W-W}$, controls the swelling and localization of the microgels. At the temperature of interest, $\epsilon_{S-S} = 0.275\varepsilon, \epsilon_{W-S} = 0.5\varepsilon$ ensures that the microgels are in a swollen state in bulk and a collapsed state after drying while also positioning the microgel at the liquid-vapor interface.

To simulate solvent evaporation and the drying of soft interfacial films onto a solid substrate, a deletion zone with a thickness of 10σ was introduced at the top of the simulation box (z=Lz). The evaporation rate was controlled by adjusting the number of vapor particles removed from the deletion zone and the frequency of their removal (details are presented in **Supporting Information**).

**Supporting Information**

Supporting Information is available from the Wiley Online Library or from the author.

**AUTHOR INFORMATION**

**Corresponding Author**

*Name: Matthias Karg

E-mail: karg@hhu.de

**ACKNOWLEDGEMENTS**

The authors acknowledge Yichu Zhou for her contributions to the synthesis of the small CS microgels in **SI**, Dr. Kiran Kaithakkal Jathavedan for providing the PNIPAM microgels in **SI**, Vahan Abgarjan for co-supervising Julian Ringling's master's project, and Dr. Marius Otten for TEM measurements, and Keno Ky Studer (ETH Zürich) and Prof. Moran Bercovici (Technion) for the fruitful discussion. M.K. and R.v.K acknowledge the German Research Foundation (DFG) for funding under grant KA3880/6-1 and KL1165/27-1 (project number 395854042) respectively.

We would like to acknowledge the Center for Advanced Imaging (CAi) at Heinrich-Heine-University Düsseldorf for providing access to the Zeiss LSM880 Airyscan system (ref. no. DFG- INST 208/746-1 FUGG). Data of this work were discussed and parts of this paper have been written at the Kavli Institute for Theoretical Physics (KITP), Santa Barbara (USA) during the "Nanoparticle Assemblies: A New Form of Matter with Classical Structure and Quantum Function" programme. This research was supported in part by the National Science Foundation under Grant No. NSF PHY-1748958. Financial support from the Deutsche Forschungsgemeinschaft (DFG) within the Collaborative Research Center "Functional Microgels and Microgel Systems" (SFB 985) and from the Russian Science Foundation, Project No. 21-73-30013, is gratefully acknowledged. The computer simulations were carried out using the equipment of the shared research facilities of HPC computing resources at Lomonosov Moscow State University.[91]

# Supporting Information

# Drying of Soft Colloidal Films

*Keumkyung Kuk,[a] Julian Ringling,[a] Kevin Gräff,[b] Sebastian Hänsch,[c] Virginia Carrasco-Fadanelli,[d] Andrey A. Rudov,[e,f] Igor I. Potemkin,[e,f] Regine von Klitzing,[b] Ivo Buttinoni,[d] and Matthias Karg[a]\**

[a]Institut für Physikalische Chemie I: Kolloide und Nanooptik, Heinrich-Heine-Universität Düsseldorf, Universitätsstr. 1, 40225 Düsseldorf, Germany

[b]Institute for Condensed Matter Physics, Soft Matter at Interfaces, Technische Universität Darmstadt, Hochschulstr. 8, 64289 Darmstadt, Germany

[c]Center for Advanced Imaging, Heinrich-Heine-Universität Düsseldorf, Universitätsstr. 1, 40225 Düsseldorf, Germany

[d]Institut für Experimentelle Physik der kondensierten Materie, Heinrich-Heine-Universität Düsseldorf, Universitätsstr. 1, 40225 Düsseldorf, Germany

[e]Leibniz Institute for Interactive Materials, 52056 Aachen, Germany

[f]Physics Department, Lomonosov Moscow State University, Moscow 119991, Russian Federation

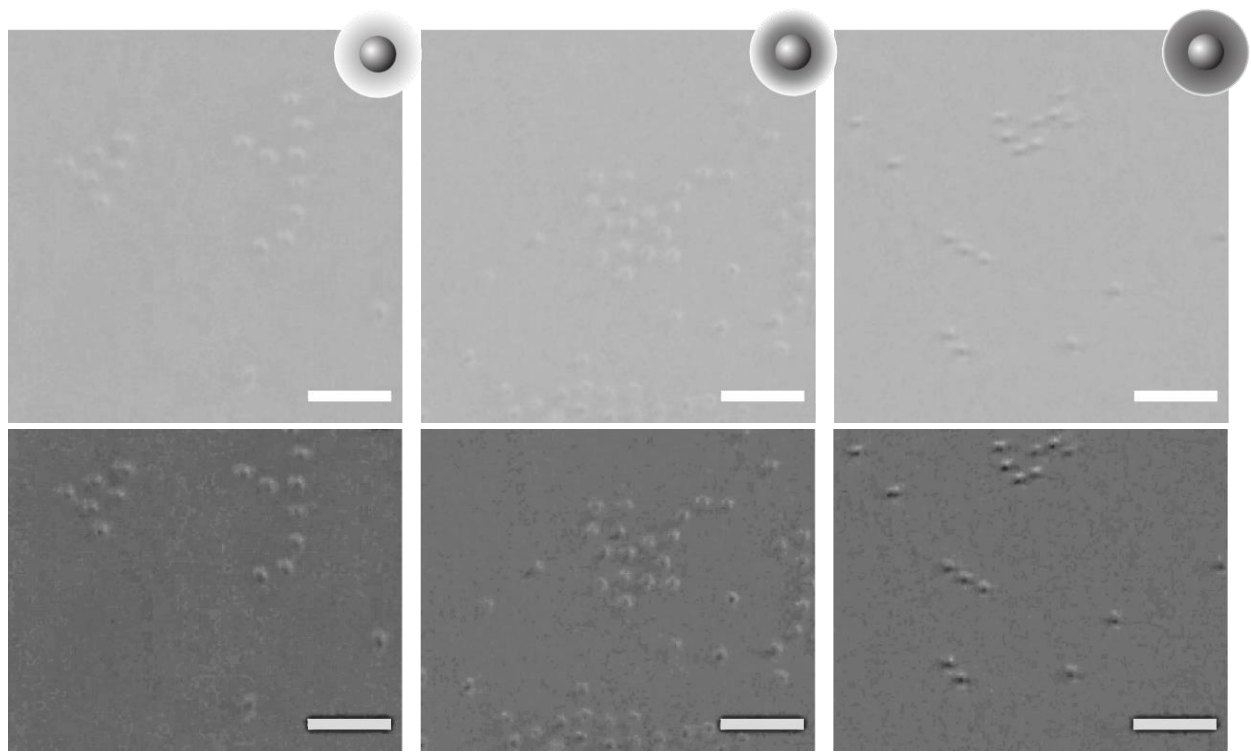

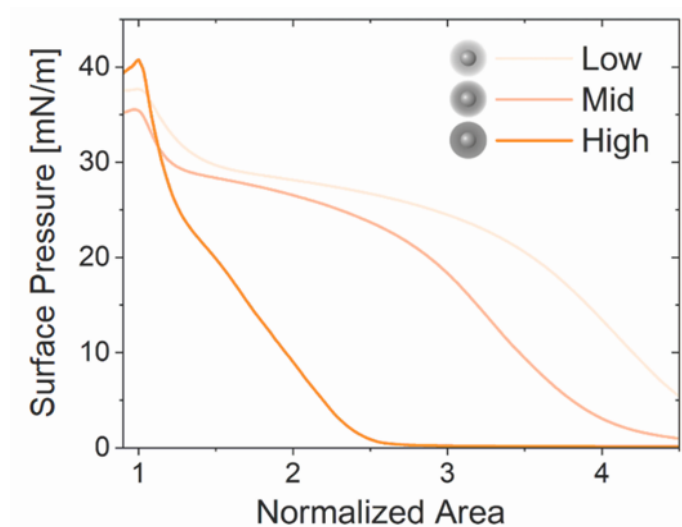

**Figure S1.** Top row: Light microscopy images (in reflection) of CS-low (left), CS-medium (middle), and CS-high (right) at the air/water interface at a surface pressure of approx. 0 mN/m. At least 200 microgels were used for the acquisition of the mean nearest center-to-center distance, $D_{c\text{-}c}$, in **Figure 1** in the main manuscript. Bottom row: Parts of the images shown in the top row with adjusted brightness and contrast to better visualize the microgels. The scale bars correspond to 5 μm. The graph below shows the compression isotherms of the corresponding microgels.

**Table S1.** The critical height, $H^*$, of CS microgels at the water/solid interface and deformation rate at the air/water interface.

|  | Water/solid interface | | | | Air/water interface | |
| --- | --- | --- | --- | --- | --- | --- |
|  | Hydrophilic | | Hydrophobic | | *Ethanolic | **Aqueous |
|  | $H^*$ [nm] | $H^*/D_h$ | $H^*$ [nm] | $H^*/D_h$ | Degree of deformation $(D_i/D_h)$ | |
| CS-low | 159 ± 35 | 15.8% | 157 ± 14 | 15.6% | 170.5% | 174.5% |
| CS-medium | 301 ± 6 | 27.0% | 169 ± 33 | 15.1% | 133.7% | 136.4% |
| CS-high | 511 ± 6 | 48.6% | 409 ± 23 | 38.8% | 109.2% | 109.2% |

*adsorption to the air/water interface from ethanolic dispersion (ethanol as the spreading agent) and **from aqueous dispersion (spontaneous adsorption). The measurements were done after 30 minutes of equilibration time for more than 200 microgels at 0 mN/m.

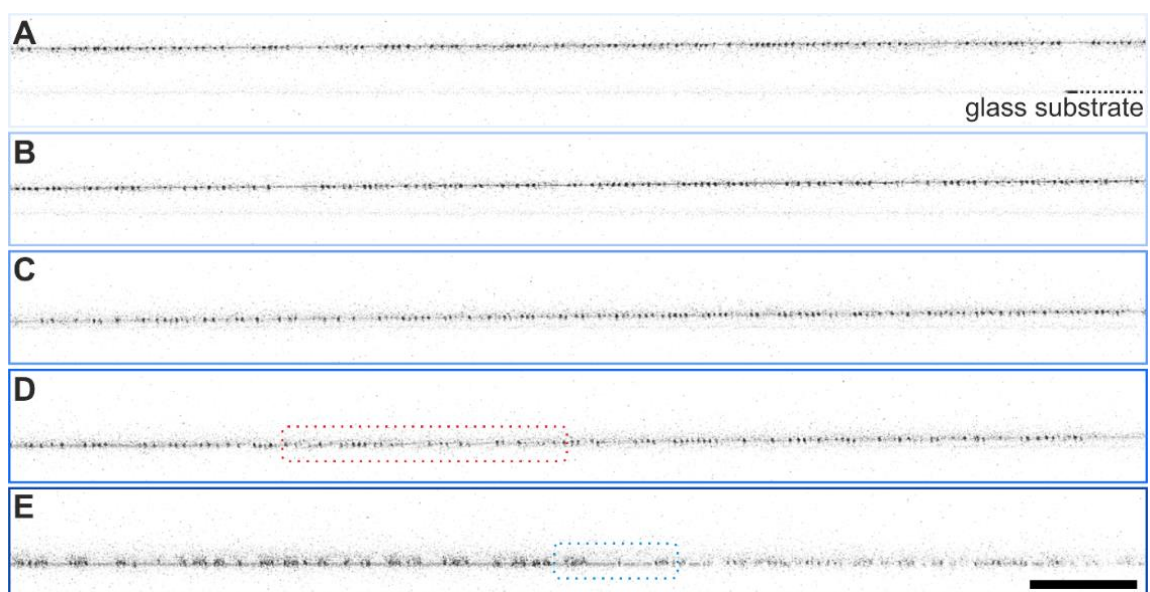

**Figure S2.** In situ monitoring of the drying CS-medium film on the hydrophilic substrates via confocal microscopy. **A-C** depict the lowering of the CS-medium monolayer (upper line, dots are rhodamine B dyed core) to the substrate (lower line). **D** shows the interference pattern that arises from the slope of the meniscus (receding wetting angles $\theta_1$, in the red box with a dotted line) and the monolayer thin film on the left. **E** shows the drying front of the microgel monolayer indicated by the higher intensity of the dried monolayer. The scale bar corresponds to 20 μm.

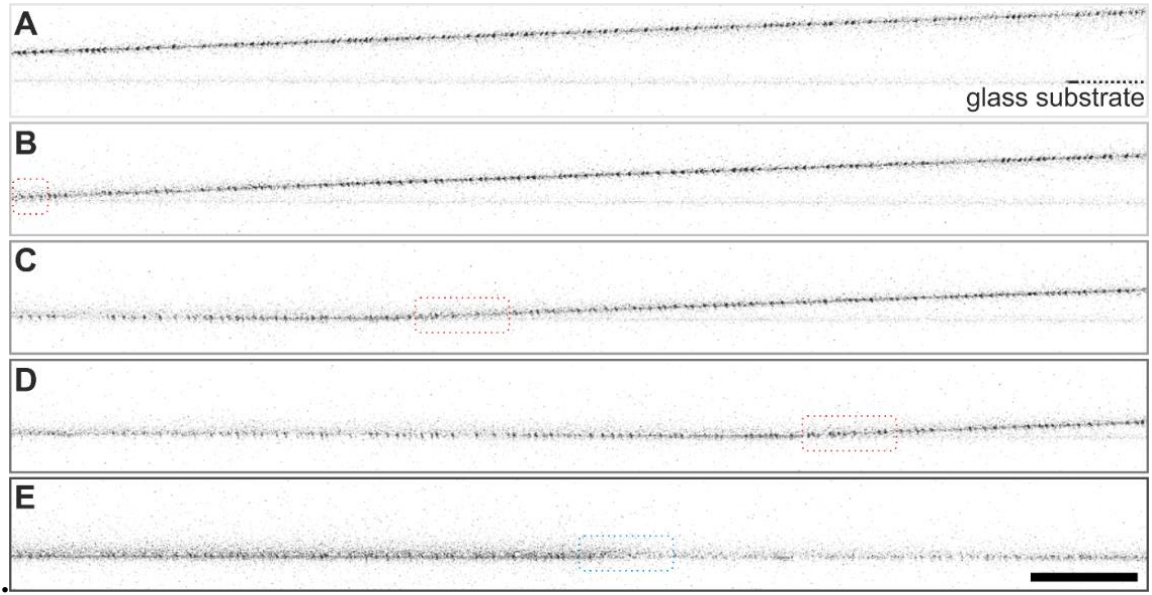

**Figure S3.** In situ monitoring of the drying CS-medium film on the hydrophobic substrate via confocal microscopy. **A** and **B** depict the lowering of the CS-medium monolayer to the substrate. **B-D** show the beginning of the monolayer thin film and the receding wetting angles, $\theta_2$. **E** shows the drying front of the microgel monolayer indicated by the higher intensity of the dried monolayer. The scale bar corresponds to 20 µm.

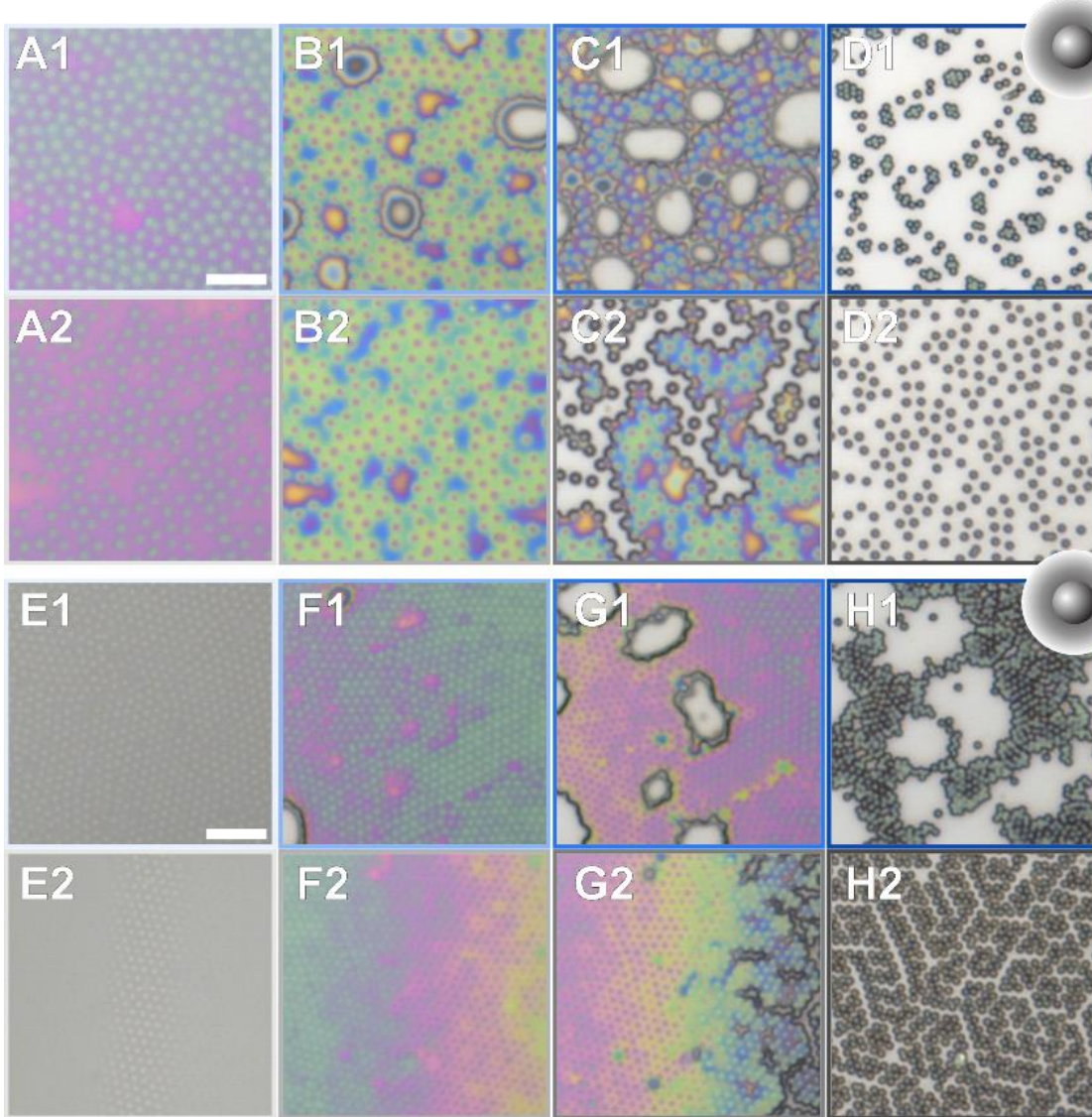

**Figure S4.** In situ monitoring of the drying of CS-medium film, transferred from Π near 10 mN/m, drying on hydrophilic (**A1**-**D1**) and on hydrophobic substrates (**A2**-**D2**). (**E1**-**H2**) are the same data sets for the microgel film transferred from Π near 30 mN/m. The scale bars correspond to 5 μm.

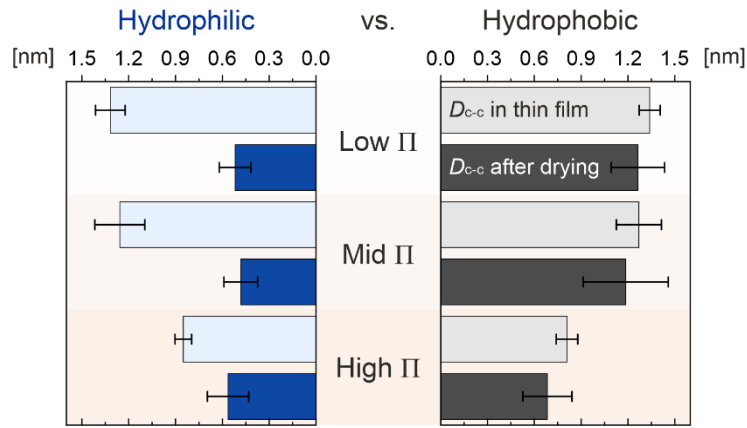

**Figure S5.** Center-to-center distance, $D_{c\text{-}c}$, of CS-medium film drying onto hydrophilic (blue) and hydrophobic (gray) substrates. Distances in thin film are shown using lighter colors and compared to the ones of fully dried microgel clusters (darker colors). The error bars are the standard deviations of the first peaks of the radial distribution functions.

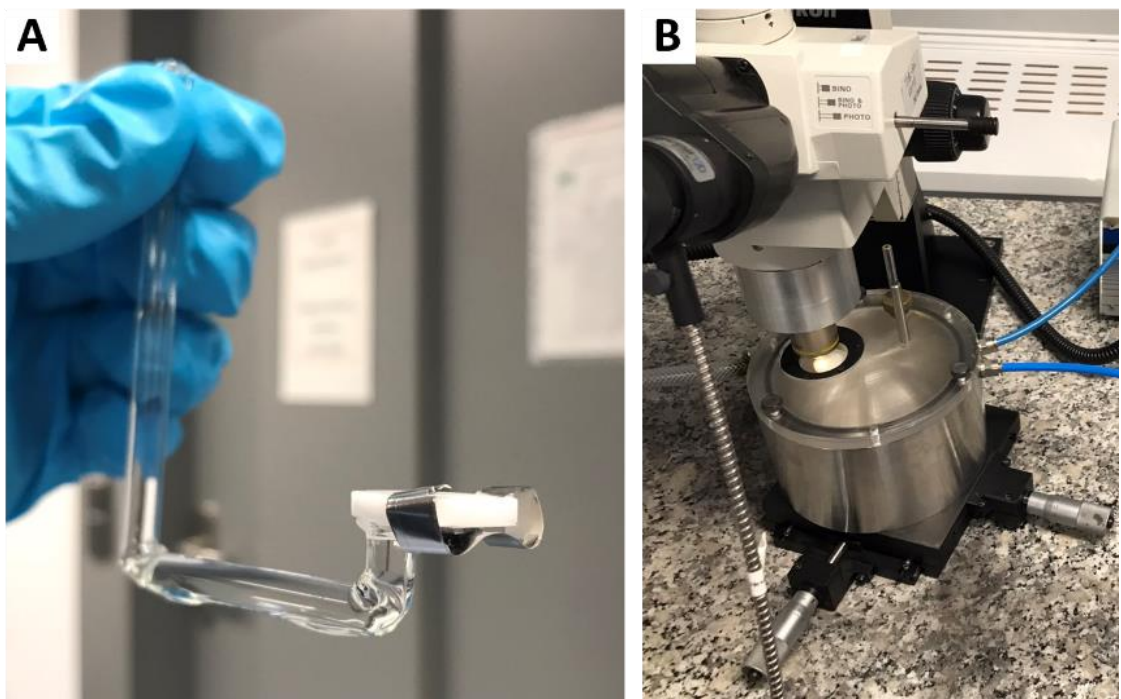

**Figure S6.** A) A porous glass plate is attached to a glass capillary tube (film holder) used for the thin film pressure balance technique. The sample is attached to the porous glass plate using a stainless-steel clamp. B) The stainless-steel pressure chamber with a quartz glass window is placed under a reflective light microscope.

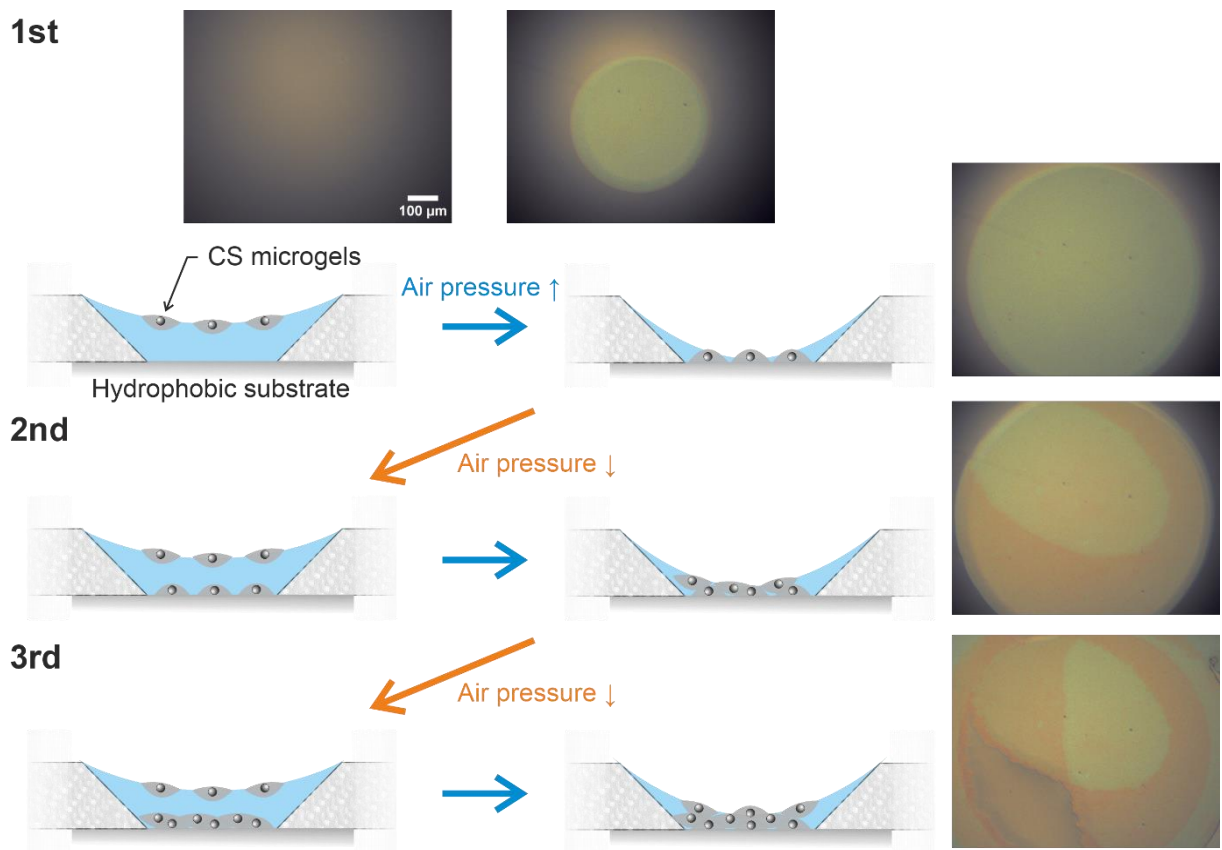

**Figure S7.** The interface oscillation experiment (3 cycles) using a modified thin film pressure balance setup, depicted in **Figure 3**, only results in the formation of multilayers (3 layers of CS-medium film, indicated by the different colors of the patches) when attempted on hydrophobic substrates.

## MODEL AND SIMULATION DETAILS

Molecular dynamics (MD) simulations were conducted using the LAMMPS package [1], employing a standard coarse-grained model with explicit solvent representation. The simulations were carried out in the NVT ensemble in reduced (dimensionless) units derived from the potential parameters, $\epsilon$ and $\sigma$, mass of a single particle, $m$, and Boltzmann's constant, $k_B$. The equations of motions were integrated with a time step, $\Delta t = 0.005\,\tau$, where $\tau = \sigma\,(m/\epsilon)^{0.5}$ is the standard time unit for a Lennard-Jones (LJ) fluid.

Three different types of coarse-grained particles, referred to as beads, are employed in the simulations: water (both liquid and vapor) beads denoted as W, solid microgel core beads designated as C, and microgel polymeric shell beads labeled as S (**Figure S8**, **Table S2**). Each beads have identical mass, m, and size $\sigma$.

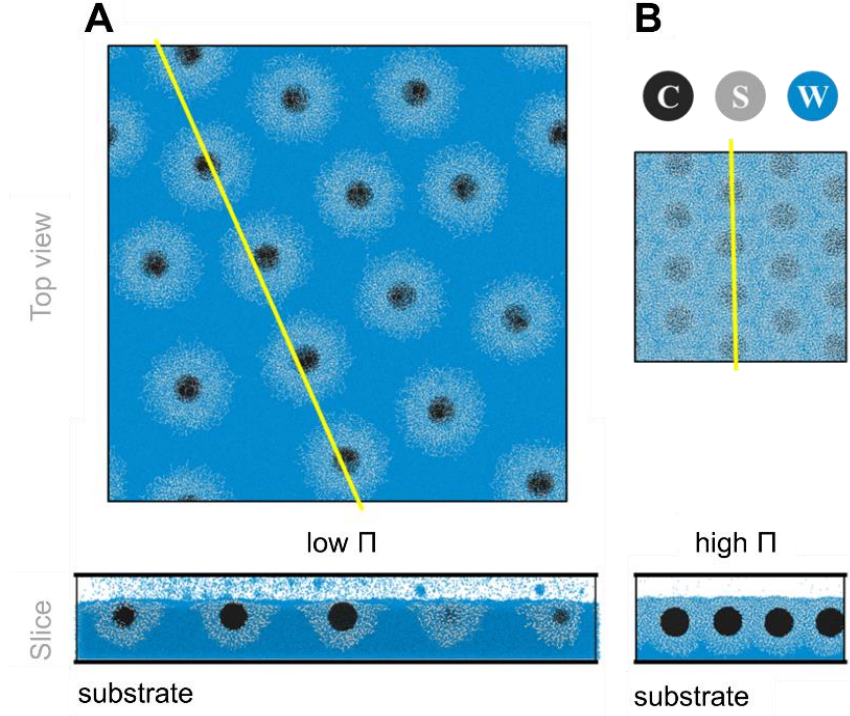

**Figure S8.** The snapshots of microgel monolayers at liquid/air interface. Cases of low (A) and high (B) compression degrees. View from the air. Solid core, C, shell, S, water, W, and beads are colored black, grey, and blue, respectively. The yellow line indicates the slice area.

Interactions within the system are governed by a standard Lennard-Jones (LJ) 12–6 pairwise potential, characterized by strength $\epsilon$ and characteristic length $\sigma$:

$$U_{LJ\,smo}(r) = \varphi_{LJ}(r) - \varphi_{LJ}(r_{cut}) \qquad \text{Eq. S1}$$

$$\varphi_{LJ}(r) = \begin{cases} 4\epsilon_{ij}\left(\left(\dfrac{\sigma}{r}\right)^{12} - \left(\dfrac{\sigma}{r}\right)^{6}\right), & r \leq r_{cut} \\ 0, & r > r_{cut} \end{cases} \qquad \text{Eq. S2}$$

where $r_{cut} = 3\sigma$, $i,j \in \{W, S, C\}$.

**Table S2:** MD interaction parameters (in units of $\epsilon$) at $T = 0.72\epsilon/k_B$ used in simulations, where $\epsilon$ refers to the LJ energy parameter of the bead-to-bead interaction and $k_B$ is the Boltzmann constant.

|       | C     | S     | W     | sub**            | cap**  |
|-------|-------|-------|-------|------------------|--------|
| C     | 0.275 | 0.275 | 2*    | 2*               | 0.001  |
| S     |       | 0.275 | 0.5   | $\epsilon_{S-sub}$ | 0.001  |
| W     |       |       | 1     | $\epsilon_{W-sub}$ | 0.001  |
| sub   |       |       |       | –                | –      |
| cap   |       |       |       |                  | –      |

*$r_c = 2^{1/6}\sigma$; **$r_c = 3\sigma$;

**Liquid–vapor coexistence**

We begin by outlining the approach taken to establish the coexistence between the liquid and vapor phases. The system encompasses W beads. The appropriate choice of LJ parameters accurately capturing both bulk and interfacial properties of simple fluids are well described in [2–6]. We set and maintained the system temperature $T = 0.72\ \epsilon/k_B$, using a Nose-Hoover thermostat. The temperature is selected within the range spanning from the triple-point temperature [7], $T = 0.65\ \epsilon/k_B$, to the critical temperature [8], $T = 1.08\ \epsilon/k_B$, thus accurately representing the coexistence of liquid and vapor phases. We defined the interaction parameter between W beads as $\epsilon_{W-W} = 1\epsilon$. Placing W beads uniformly distributed over the empty simulation box eventually leads to the formation of a liquid aqueous phase(droplet) surrounded by water vapor (**Figure S9**). The liquid phase is characterized by a density of $\rho \sim 0.8/\sigma^3$. The vapor is in dynamic equilibrium with a liquid. Droplet size becomes constant and does not change over time due to the dynamic balance between condensation and evaporation of the W beads.

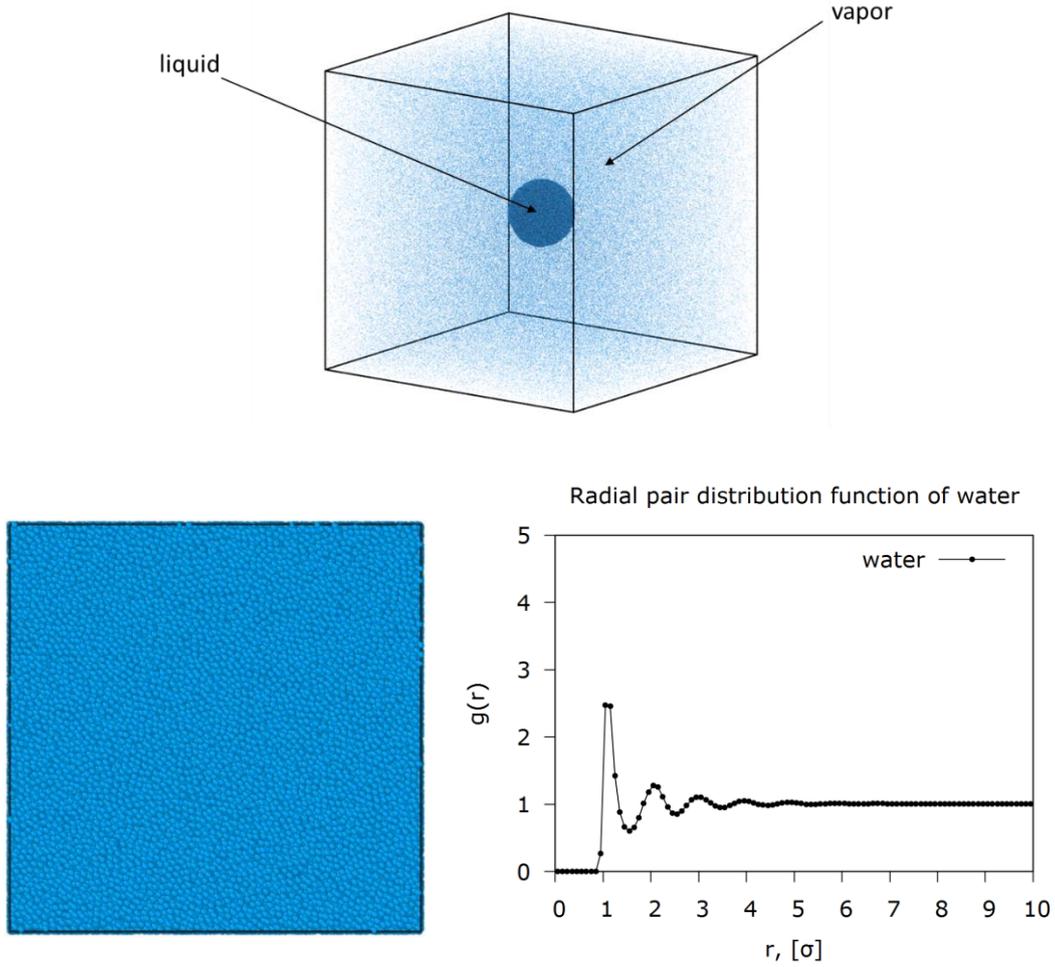

**Figure S9:** (Top) Liquid-vapor coexistence. (Left) Schematic illustration of the liquid phase. Liquid density $\rho = 0.8/\sigma^3$. (Right) Radial pair distributions function of beads in a liquid phase at $T = 0{,}72\epsilon/k_B$.

**Substrate definition**

To establish a substrate, we implemented a simplified smooth solid flat wall (sub) represented by an LJ potential (Eq. S2) positioned at the bottom of the simulation box, specifically at z = 0, and functioning as the substrate. This wall exerts a force on each bead in the system in the z-direction, perpendicular to the wall, whenever the distance between the bead and the wall is less than $r_{cut\ wall} = 3\sigma$. Similarly, on the opposite side, at $z = L_z$, we introduced a second wall, serving a technical auxiliary purpose. All interaction parameters between the upper wall and every bead in the system were selected as $\epsilon_{i-top} = 0.001\varepsilon$, $r_c = 2^{1/6}\sigma$, $i \in \{W, C, S\}$ representing complete repulsion. Consequently, we achieved a 2D+1 slab geometry, with periodic boundary conditions exclusively applied in the x and y directions.

This simplification of the substrate description is justified, considering that the roughness of the glass surface and the PFOES coating are much smaller than the size of the microgel. In our study, we focused solely on the effect of surface hydrophobicity/hydrophilicity at a given temperature. By varying the water-substrate interaction parameter, $\epsilon_{W-sub}$, and microgel-interaction parameter, $\epsilon_{S-sub}$, we could control the substrate's affinity for the liquid and polymer beads, thereby altering the contact angle of the droplet $\theta$ (**Figure S10**). It is noteworthy that the intrinsic contact angle $\theta$ generally depends on the roughness and density of the substrate [9,10], $\rho_s$, the LJ energy parameter of the substrate−droplet interaction and the temperature T of the system [11].

To reproduce the specific hydrophobicity/hydrophilicity of the substrate, we determined the intrinsic contact angle $\theta$ of a droplet of pure liquid as a function of the interaction strength parameter $\epsilon_{W-sub}$. We placed a spherical water droplet beside the substrate and monitored the shape change. It was found that $\epsilon_{W-sub} = 1.7\epsilon$ and $3.8\epsilon$ correspond to intrinsic contact angles of approximately $\theta \sim 100°$ and $\sim 3°$ for hydrophobic and hydrophilic substrates, respectively. By adjusting the microgel-substrate interaction parameter, $\epsilon_{S-sub}$, we could regulate the substrate's affinity for the microgels. We explore two distinct shell/substrate interaction scenarios: one with a strong adhesion ($\epsilon_{S-sub} = 20\epsilon \gg \epsilon_{W-sub}$), resulting in significant microgel spreading and the other with a weaker interaction ($\epsilon_{S-sub} = 3\epsilon \sim \epsilon_{W-sub}$), leading to partial wetting.

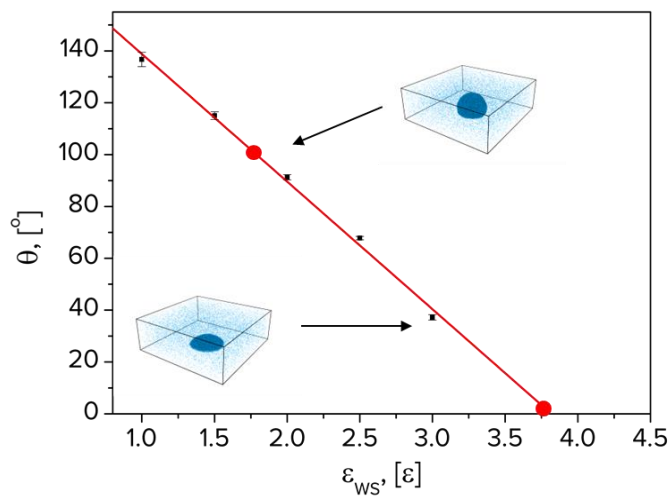

**Figure S10:** Dependence of contact angle of a water droplet as a function of the interaction energy $\epsilon_{W-sub}$ between water and substrate obtained in the computer simulation. For $\epsilon_{W-sub}$ values of $1.7\epsilon$ and $3.8\epsilon$ were chosen to obtain the intrinsic contact angles of the pure water at hydrophobic and hydrophilic substrates, respectively.

**Microgel overview: model, parametrization, characteristics**

CS microgel featuring a hydrodynamic diameter of 40±0.6σ with fuzzy polymeric shells and a spherical solid core of diameter 16σ has been created (**Table S3**). The average fraction of cross-linkers in the polymeric shells equals 9.4%. The core-shell structure was designed following the methodology outlined in our prior works. Initially, a unit cell of the diamond crystal lattice was created, with tetrafunctional cross-linkers at the vertices. Subsequently, two cubic supercells denoted as S50 and S25, were constructed. S50, comprised of 50×50×50 unit cells, serves as the template for the solid nanoparticle, while S25, consisting of 25×25×25 unit cells, acts as the template for the polymeric shell.

A solid nanoparticle was constructed by inscribing a spherical frame into an S50 supercell and removing all beads outside the sphere. The beads composing the nanoparticle are labeled as C.

**Table S3:** Characteristics of the core-shell microgel with an anisotropic distribution of cross-links.

| Sample | Core | Inner shell | | | | Outer shell | | | | Shell(Total) | |
|---|---|---|---|---|---|---|---|---|---|---|---|
| | $N_{beads}$ | $N_{beads}$ | $N_{cross-links}$ | $\bar{l}$ | $\sigma_l$ | $N_{beads}$ | $N_{cross-links}$ | $\bar{l}$ | $\sigma_l$ | $N_{total}$ | $\overline{\%}$ cross-links |
| CS | 3020 | 4342 | 778 | 3 | 1 | 12654 | 810 | 6 | 1 | 16996 | 9.4 |

In line with experimental findings, a simulation was conducted to replicate the structure of core-shell microgels, featuring an inhomogeneous cross-link distribution across their inner and outer regions. This was achieved by constructing a polymeric shell around the solid nanoparticle using a scaled S25 supercell. Within this shell, all bonds between tetrafunctional atoms were substituted with polymer subchains of varying lengths. The length of these subchains dictated the degree of cross-linking within the microgel: shorter subchains corresponded to higher cross-link densities, and vice versa. Different laws can describe the distribution of the chain lengths. We will use a symmetric Gaussian distribution with an average value of $\bar{l}$, and a standard deviation of $\sigma_l$. $\sigma_l = 0$ corresponds to the case of an ideal diamond-like network used in many simulations. The higher the $\sigma_l$ the stronger the discrepancy of the network from the ideal regular structure. Then, three spherical frames of different radii were inscribed into the modified supercell. A small frame is necessary for forming the void in the polymeric network of the same size as the solid nanoparticle. The sizes of the middle, $R_M$, and large, $R_L$, frames control the thickness of the polymeric shell's inner and outer regions, respectively. All beads were cropped inside the small and outside of the large ellipsoidal frames. The rest of the beads forming the microgel shell were denoted as S. Pair of ($\bar{l}_{in}$, $\sigma_{in}$) and ($\bar{l}_{out}$, $\sigma_{out}$) set the distribution of the cross-linkers within the regions. Such a procedure allows us to synthesize polymeric shells with highly cross-linked inner parts and gradually decrease cross-linker concentration towards the outer part. The solid nanoparticle was inserted into the void of the microgel with further grafting of the dangling chains of the polymeric shell to the nanoparticle surface. The dangling chains of the polymeric shell were physically attached to the nanoparticle surface. Going through all of the free ends of the dangling chains, we monitor whether the distance between it and the closest bead at the surface of the solid core satisfies the condition: $|r_{end} − r_{surf}| \leq r_c$. If so, a bond between the free end of the dangling chains and the bead of the surface of the solid core was formed. Only one bond could be formed between the free end and the solid core.

Connectivity of the beads into a polymer network was maintained by a combination of the finite extension nonlinear elastic (FENE) potential and Lennard-Jones potential:

$$U_{bond}(r) = U_{FENE}(r) + U_{LJ}(r) \qquad \text{Eq. S3}$$

where the distance between two beads is denoted by r.

$$U_{FENE}(r) = -\frac{1}{2}KR_{max}^2 \ln\left(1 - \frac{r^2}{R_{max}^2}\right) \qquad \text{Eq. S4}$$

$$U_{LJ}(r) = 4\epsilon_{bond}\left[\left(\frac{\sigma}{r}\right)^{12} - \left(\frac{\sigma}{r}\right)^6\right] + \epsilon_{bond} \qquad \text{Eq. S5}$$

with the spring constant, $K = 15\ \epsilon/\sigma^2$, the maximum bond length, $R_{max} = 1.5\ \sigma$, $\epsilon_{bond} = 1\epsilon$, and the cutoff radius $r_{cut} = 2^{1/6}\ \sigma$.

The angle potential, which is necessary to maintain the spherical shape of the solid core and prevent it from deformation at the interfaces, is given by

$$U_{angle} = K_a(\theta - \theta_0), \qquad \text{Eq. S6}$$

where $K_a = 100\epsilon$ is the bending stiffness, and $\theta_0 = 109.47°$ is the angle between two pairs of connected C beads sharing a common bead.

To mimic the swollen state of the microgel in the bulk solution, we set the value of the LJ parameter for the microgel bead-to-bead interactions to $\epsilon_{S-S} = 0.275\epsilon$. For the interactions between the shell and water, $\epsilon_{S-W} = 0.5\epsilon$ was chosen. The idea behind these is as follows: at $\epsilon_{S-W} = 0.5\epsilon$ and $\epsilon_{S-S} = 0.275$ all subchains of the microgel in the bulk are elongated, providing maximum swelling. At the same time, in the absence of solvent (after the complete evaporation of the liquid), the polymer shell collapses. The values for the core-water interaction $\epsilon_{C-W} = 2\epsilon, r_c = 2^{1/6}\sigma$ are chosen to be purely repulsive to prevent flocculation and ensure the insolubility of particles.

**Microgel monolayer formation at the air/liquid interface**

A planar liquid/vapor interface was established and aligned parallel to the xy plane. In proximity to the interface, 16 microgels with identical architectures were randomly distributed. The degree of compression was controlled by concurrently adjusting the dimensions of the simulation box in both the X and Y directions. For low $\Pi$ values, the simulation cell was configured as a rectangular box with dimensions $L_x = L_y = 260\sigma, L_z = 50\sigma$. Conversely, for high $\Pi$ values, the dimensions were adjusted to $L_x = L_y = 120\sigma, L_z = 50\sigma$. **Figure S8** provides an overview of the monolayers of CS microgels at two different compression states – low (A) and high $\Pi$ (B). Similar to [12], the microgel at the interface is a non-flat object. It consists of a three-dimensional swollen part, namely the bulky part (B-part), immersed in the liquid phase, and a thin two-dimensional polymeric layer, namely the substrate part (S-part), at the interface. At low $\Pi$ (**Figure S8A**), $L_x = L_y = 260\sigma$, some of the microgels start to come into contact with each other. According to [12] the microgel surface coverage is attributed to the region of the first growth on the compression curve, where the 2D S-parts of the microgels start to be in contact with each other. We performed a qualitative analysis of the monolayer. We calculated the center of mass of each particle and computed the average 2D radial distribution functions, g(r). Examples of the 2D Voronoi tessellation of the simulation box and representative g(r) are shown in **Figure S11**. The position of their first maximum of g(r) allows us to estimate inter-particle distance, $D_{c\text{-}c} = 50.1 \pm 0.4\sigma$. (**Table S4**). At $L_x = L_y = 120\sigma$, the surface pressure of the monolayers proceeds to the second increase (high $\Pi$ value). The 2D shell-to-shell contacts are replaced by 3D interactions of the B-part of the microgels (**Figure S8B**). $D_{c\text{-}c} = 31.2 \pm 0.2\sigma$.

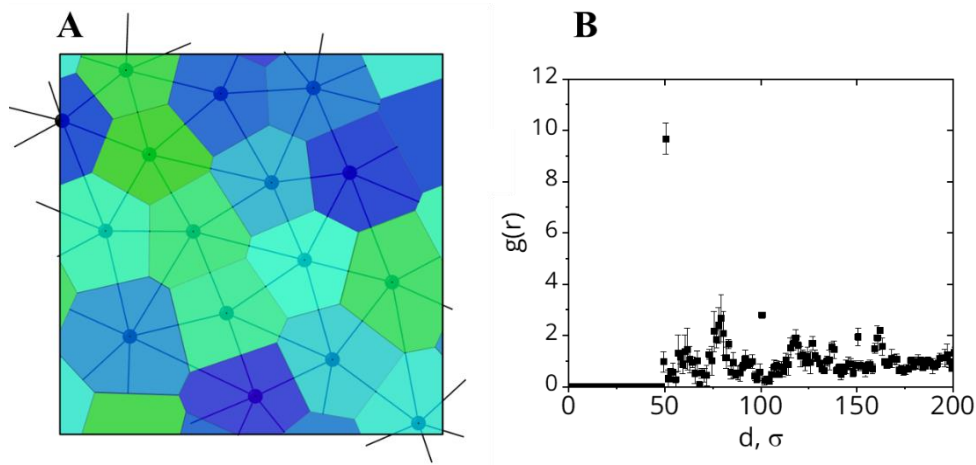

**Figure S11.** (A) 2D Voronoi tessellation of the simulation box, taking the center of mass of the solid core particles as Voronoi cell centers. (B) The radial pair distribution function, RDF, is for the center of mass of the solid core particles. Cases of low compression degree. Low surface affinity to microgels.

**Drying dynamics and structural changes of microgel monolayer**

We aim to compare monolayers transferred directly onto both hydrophilic and hydrophobic substrates. Following an equilibration run of $10^7$ timesteps at the air/liquid interface, we initiate evaporation (**Figure S12A**). This involves introducing a deletion zone with a thickness of $10\sigma$ at the top of the simulation box, positioned at $z = L_z$. Within this zone, water vapor beads are consistently removed from the system to modulate the balance between condensation and evaporation. The evaporation rate is regulated by the number of particles removed and the frequency of removal. Specifically, every $0.5\tau$, all solvent beads were extracted, emulating rapid solvent evaporation in a vacuum. This setup allows us to explore the scenario wherein monolayer adsorption proceeds concurrently with evaporation.

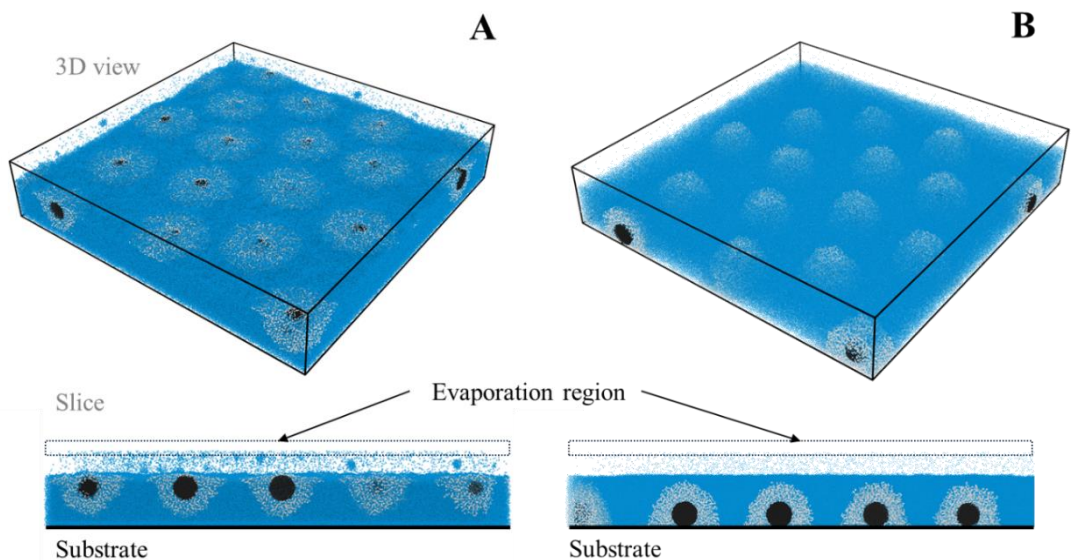

**Figure S12:** (A) Equilibrium structure of monolayer of microgels near the solvent/air interface. (B) The microgels' monolayer structure is equilibrated in a liquid phase near the substrate.

**Figure 5** presents the time evolution of the monolayers (obtained at low Π) during drying. Utilizing an orange-to-pink color gradient, we visualize variations in water layer height across distinct film regions. This dynamic phenomenon encompasses solvent redistribution, microgel shape and size changes, and morphological shifts within the monolayer.

It is noteworthy that the evaporation process within the monolayer exhibits non-uniformity. Notably, rapid evaporation occurs predominantly in the spaces between the microgels, where the absence of the polymeric network facilitates efficient vapor escape. Conversely, the network-like structure of the microgels restricts evaporation within their vicinity. This inherent asymmetry initiates the development of a drying front within the polymeric-free voids, prompting the displacement of adjacent microgels toward the drying direction. Consequently, we observe a discernible migration of the microgels from their initial positions. To regulate the affinity of the microgels to the substrate, we varied the parameter $\epsilon_{S-sub}$ within the range [$3\epsilon$, $20\epsilon$]. A higher $\epsilon_{S-sub}$, corresponds to a stronger attraction between the substrate and the microgels. It is evident that even at $\epsilon_{S-sub} = 20\epsilon$ the microgel-to-substrate adhesion is insufficient to prevent XY displacement. Nevertheless, the mobility of the microgels on the hydrophobic substrate is impeded. The higher the affinity, the less susceptible the microgels are to displacement. This is corroborated by a higher $D_{c-c}$, value and the distribution of the microgels in the dried layer, as detailed in **Table S4**.

In the context of solvent evaporation, microgels undergo significant shape and size alterations. As they dry, microgels tend to collapse, albeit with varying degrees of spreading. Microgel-substrate contacts are advantageous on hydrophobic surfaces, prompting microgels to displace solvent underneath, thereby increasing contact area with the substrate (**Figure 4C**), leading to microgel spreading. Conversely, hydrophilic surfaces favor contact with water, resulting in a minimal contact area with the substrate (**Figure 4B**). This interplay culminates in the spontaneous displacement of microgels and the disruption of the observed monolayer morphology at the water/air interface under the given surface pressure ($\Pi$).

**Table S4**. Interparticle distance, $D_{c-c}$, of dried CS monolayers on hydrophilic substrates and hydrophobic substrates in a thin film, as compared to that at Air/water interphase. $\epsilon_{S-sub}$ is the affinity of the microgels to the substrate. $\epsilon_{S-sub} = 20\epsilon$ corresponds to the case of high affinity, while $\epsilon_{S-sub} = 3\epsilon$ – low one.

|  | Low $\Pi$ | | High $\Pi$ | |
| --- | --- | --- | --- | --- |
|  | hydrophilic $D_{c-c}$, [$\sigma$] | hydrophobic $D_{c-c}$, [$\sigma$] | hydrophilic $D_{c-c}$, [$\sigma$] | hydrophobic $D_{c-c}$, [$\sigma$] |
| $\epsilon_{S-sub} = 20\epsilon$ | 36.2± 0.3 | 38.1± 0.2 | 29.7± 0.2 | 29.8± 0.2 |
| $\epsilon_{S-sub} = 10\epsilon$ | 34.4± 0.2 | 35.6± 0.2 | 29.7± 0.3 | 29.6± 0.2 |
| $\epsilon_{S-sub} = 3\epsilon$ | 28.3± 0.2 | 28.9± 0.2 | 29.8± 0.3 | 29.7± 0.3 |
| Air/water | 50.1± 0.4 | | 31.2± 0.2 | |

In the case of the film obtained under high compression ($\Pi$), as the drying process ensues, the film thickness diminishes, accompanied by a slight collapse of the microgels, as depicted in **Figure S13,14**. $D_{c-c}$ value fluctuates around 29.7$\sigma$, a measure akin to the $D_{c-c}$ of the dry film at low $\Pi$, demonstrating good agreement with experimental observations. In contrast, on the hydrophilic surface, we did not observe any deviation from the initial position of the microgels. This lack of deviation can be attributed to the influence of periodic boundary conditions with a constant number of gels, ensuring system closure. It is crucial to note that in a real experiment, the system is open after the gels are transferred to the substrate, leading to a variation in the number of gels per surface unit. Due to their already collapsed state and minimal surface interactions, the microgels demonstrate heightened mobility, facilitating facile migration.

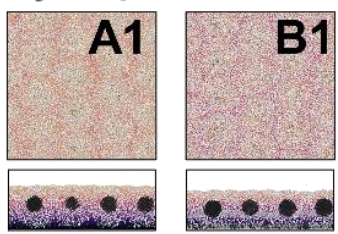
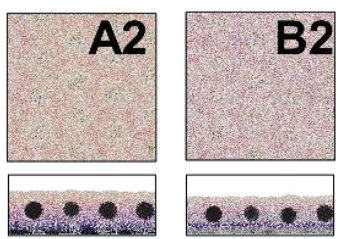

**Figure S13.** Illustration (top views) of the different stages of microgel film obtained at high Π values drying on hydrophilic (A1,B1) and hydrophobic (A2,B2) substrates as revealed by computer simulations. The narrow panels below the top views are the corresponding side views, i.e., cross-sections of each panel along Z. The height of the film with residual water is depicted by the vertical color bar. White regions in the snapshots correspond to state after water evaporation.

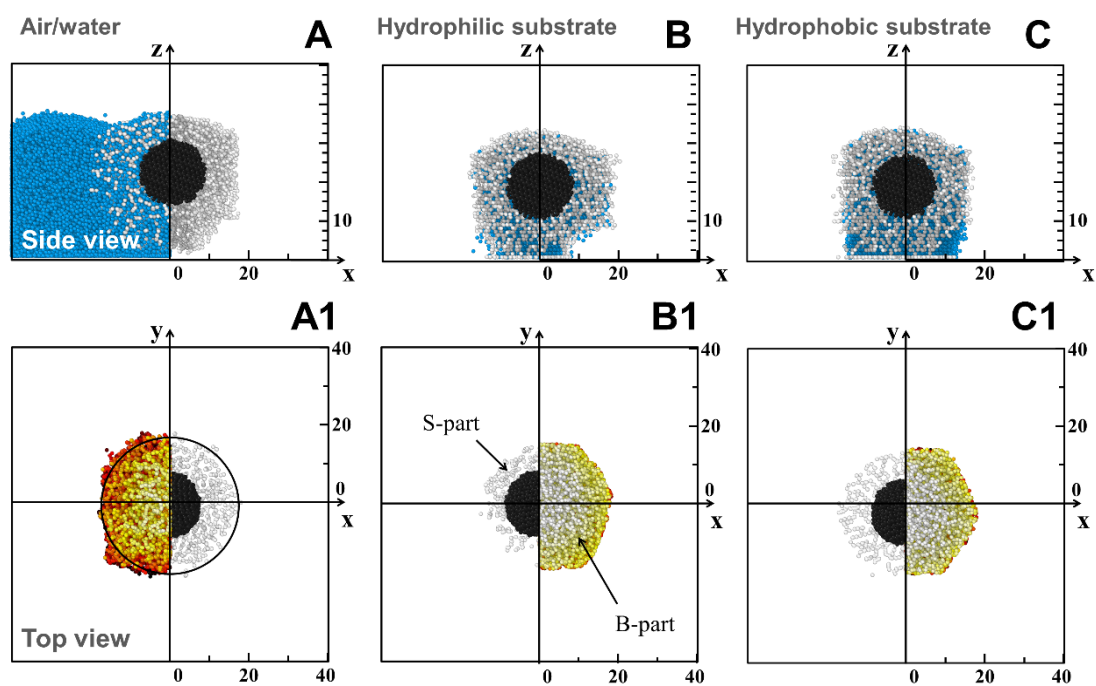

**Figure S14.** Illustration of the allocated microgel structure in the monolayer before (A-A1) and after drying in the case of hydrophilic (B-B1) and hydrophobic (A2-E2) substrates. The monolayer was transferred at a high compression degree. (A1-C1) Height map / Contact area at the interface. All snapshots are obtained using the Open Visualization Tool (OVITO) [13].

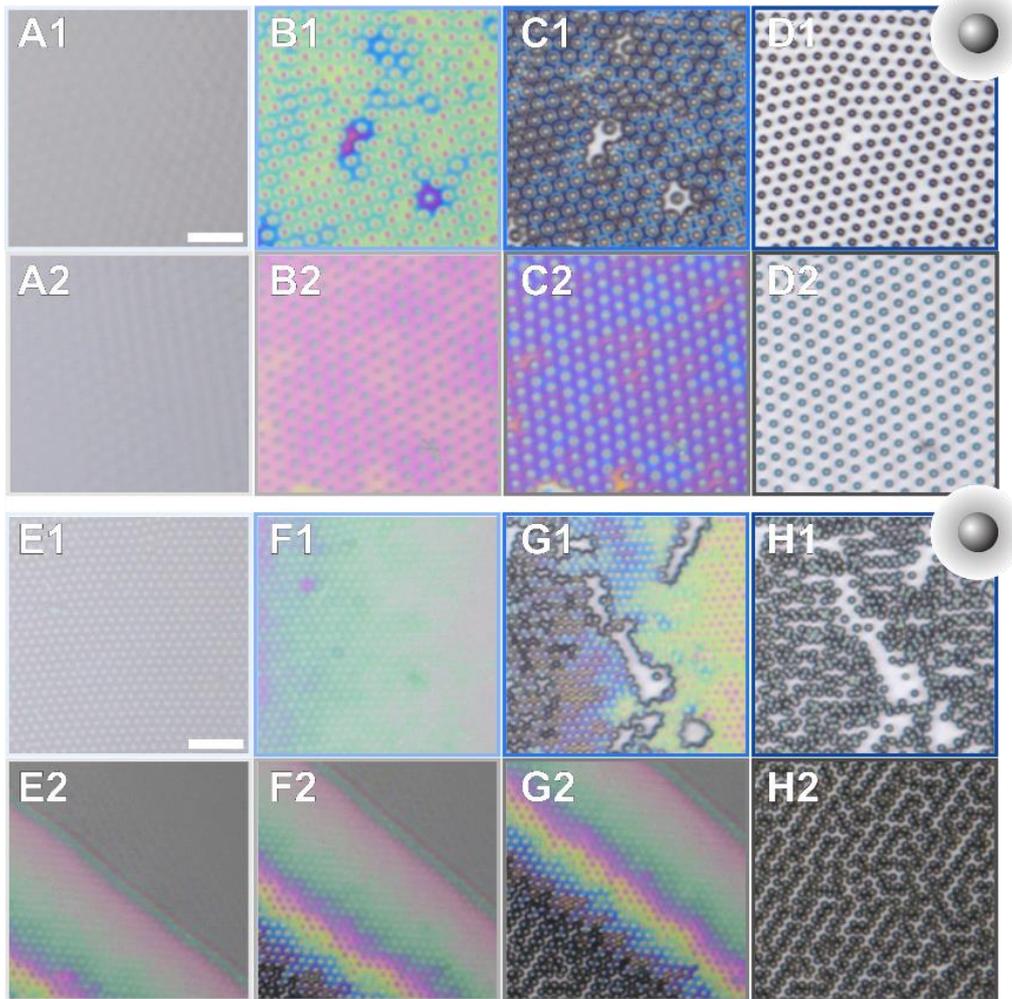

**Figure S15.** In situ monitoring of the monolayer of CS-low transferred from Π near 10 mN/m, drying on hydrophilic (**A1**-**D1**) and on hydrophobic substrates (**A2**-**D2**). (**E1**-**H2**) are the same sets of data for the monolayer transferred from Π near 30 mN/m. The scale bars correspond to 5 μm.

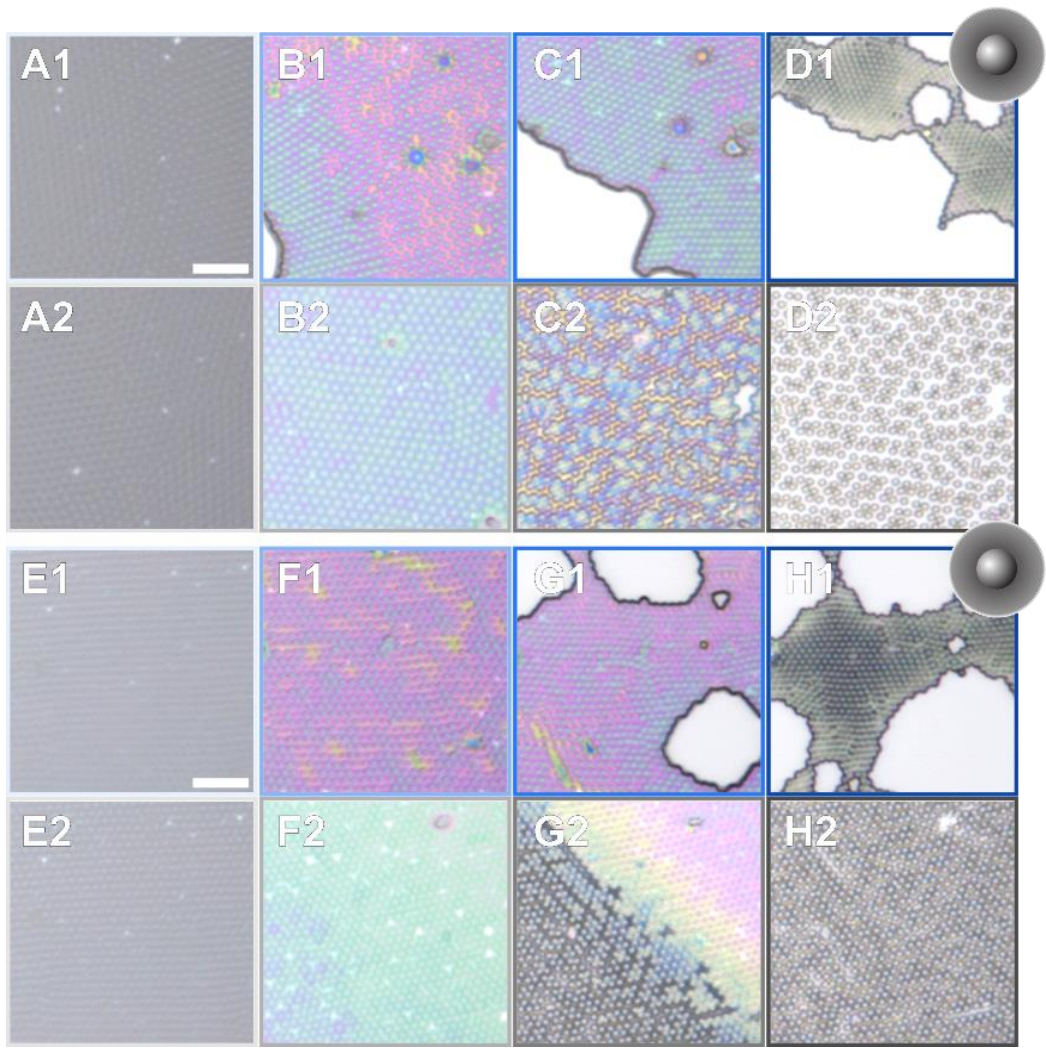

**Figure S16.** In situ monitoring of the monolayer of CS-high transferred from Π near 10 mN/m, drying on hydrophilic (**A1**-**D1**) and on hydrophobic substrates (**A2**-**D2**). (**E1**-**H2**) are the same sets of data for the monolayer transferred from Π near 30 mN/m. The scale bars correspond to 5 μm.

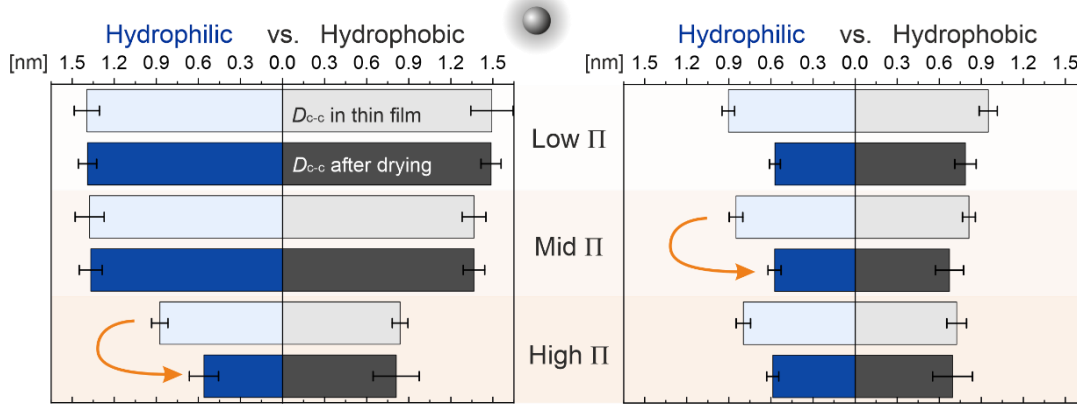

**Figure S17**. Center-to-center distance, $D_{c\text{-}c}$, of CS-low and CS-high. Films drying on hydrophilic (blue) and hydrophobic substrates (gray) in the thin fluid film (lighter color), as compared to that of dried monolayer (darker color).

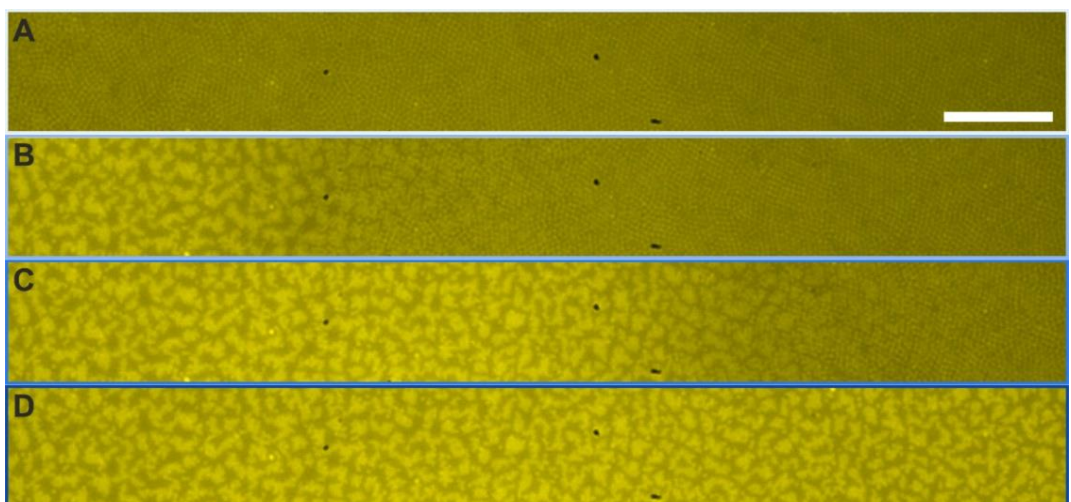

**Figure S18.** Drying of coreless PNIPAM microgel (sMG, $D_h \approx 800$ nm) film on a hydrophilic substrate via fluorescence microscopy, scale bar corresponds to 20 μm.

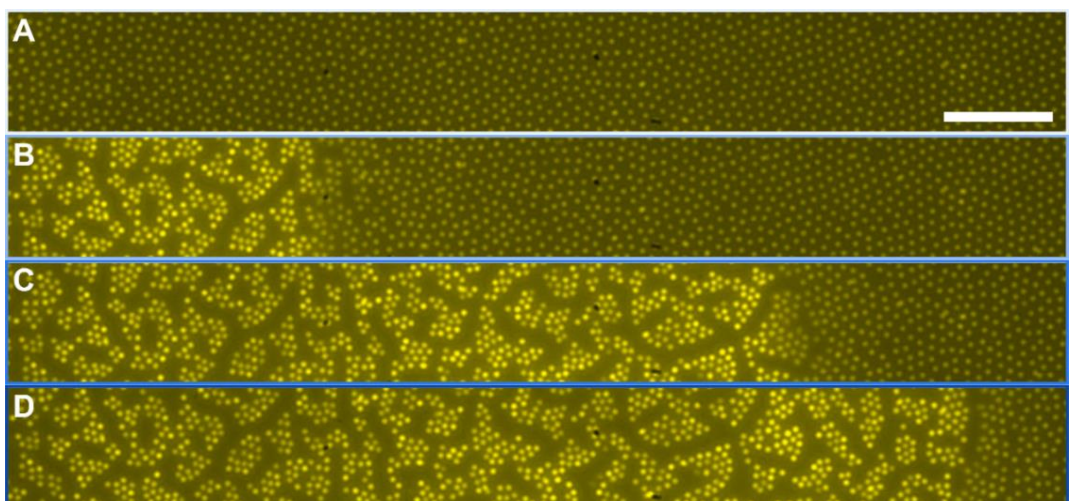

**Figure S19.** Drying of coreless PNIPAM microgel (lMG, $D_h \approx 1.5$ μm) film on a hydrophilic substrate via fluorescence microscopy, scale bar corresponds to 20 μm.

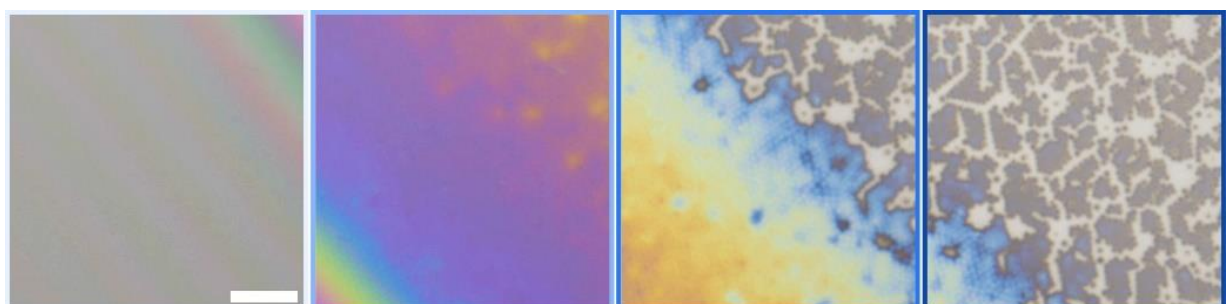

**Figure S20.** Drying of a small CS microgel ($D_c$: 105 nm $D_h$: 504 nm, cross-linker density: 5 mol.%) film on a hydrophilic substrate, transferred from Π approx. 30 mN/m, scale bar corresponds to 5 μm.

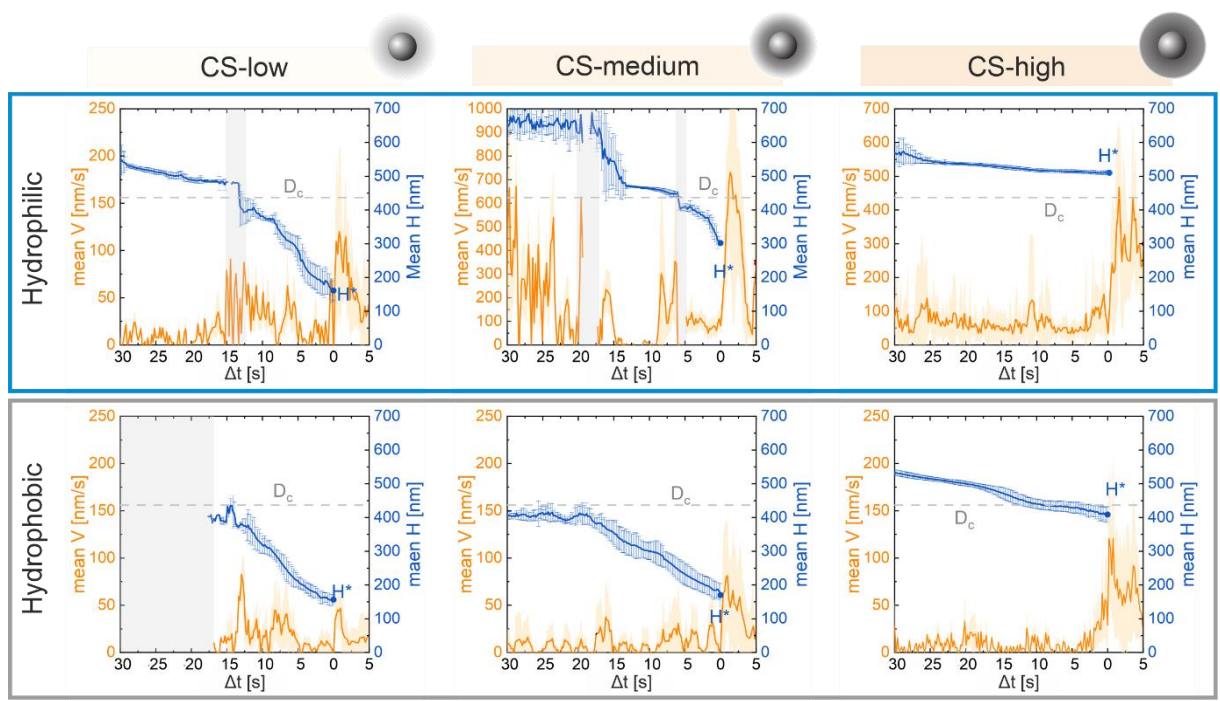

**Figure S21**. The mean velocity, *V*, and film height, *H*, of 10 randomly chosen CS microgels are plotted against normalized time, where 0 is the onset of the collapse of the individual microgels. $D_c$ and $H^*$ denote the diameter of the core and the critical height of the monolayer, respectively. The shadowed areas (grey) are frames where the monolayers were out-of-focus. Monolayer drift with preferred directions is baseline-subtracted.